\documentclass[useAMS,usenatbib,letterpaper]{mn2e}
\usepackage[totalwidth=480pt,totalheight=680pt]{geometry}
\usepackage{graphicx,amssymb,amsbsy}
\input psfig

\newcommand{\be}{\begin{equation}}
\newcommand{\ee}{\end{equation}}
\def\lta{\,\raise 0.3 ex\hbox{$ < $}\kern -0.75 em
 \lower 0.7 ex\hbox{$\sim$}\,}
\def\gta{\,\raise 0.3 ex\hbox{$ > $}\kern -0.75 em
 \lower 0.7 ex\hbox{$\sim$}\,} 
\newcommand{\cross}{\langle \sigma \rangle} 
\newcommand{\vbar}{\langle v \rangle} 
\newcommand{\nbar}{\langle n_\ast \rangle} 
\newcommand{\amax}{{a_{\rm max}}} 
\newcommand{\eb}{e_{\rm b}} 
\newcommand{\mone}{M_{1\ast}} 
\newcommand{\mtwo}{M_{2\ast}} 
\newcommand{\phase}{\theta_{\rm b}} 
\newcommand{\vdisp}{v_{\rm b}}

\title[Cross Sections for Planetary Systems Interacting with Passing Stars and Binaries] 
{Cross Sections for Planetary Systems \\
Interacting with Passing Stars and Binaries} 
\author[Li \& Adams]{Gongjie Li$^1$ and Fred C. Adams$^{2,3}$\\
$\,$ \\ 
$^1$Harvard-Smithsonian Center for Astrophysics, Institute for Theory and Computation, 
60 Garden Street, Cambridge, MA 02138\\ 
$^2$Physics Department, University of Michigan, Ann Arbor, MI 48109\\
$^3$Astronomy Department, University of Michigan, Ann Arbor, MI 48109} 

\begin{document} 

\date{submitted August 2014} 

\pagerange{\pageref{firstpage}--\pageref{lastpage}} \pubyear{2014}

\maketitle

\label{firstpage}

\begin{abstract} 
Most planetary systems are formed within stellar clusters, and these
environments can shape their properties. This paper considers
scattering encounters between solar systems and passing cluster
members, and calculates the corresponding interaction cross sections.
The target solar systems are generally assumed to have four giant
planets, with a variety of starting states, including circular orbits
with the semimajor axes of our planets, a more compact configuration,
an ultra-compact state with multiple mean motion resonances, and
systems with massive planets. We then consider the effects of varying
the cluster velocity dispersion, the relative importance of binaries
versus single stars, different stellar host masses, and finite
starting eccentricities of the planetary orbits. For each state of the
initial system, we perform an ensemble of numerical scattering
experiments and determine the cross sections for eccentricity
increase, inclination angle increase, planet ejection, and capture.
This paper reports results from over 2 million individual scattering
simulations.  Using supporting analytic considerations, and fitting
functions to the numerical results, we find a universal formula that
gives the cross sections as a function of stellar host mass, cluster
velocity dispersion, starting planetary orbital radius, and final
eccentricity. The resulting cross sections can be used in a wide
variety of applications. As one example, we revisit constraints on the
birth aggregate of our Solar System due to dynamical scattering and
find $N\lta10^4$ (consistent with previous estimates).
\end{abstract} 


\section{Introduction} 
\label{sec:intro} 

A large fraction of planetary systems form within stellar clusters
\citep{ladalada,porras} and these birth environments can influence
their resulting properties (e.g., see the reviews of
\citealt{adams2010,pfalzner}). One potentially important process
occurs when binary systems --- and single stars --- fly past solar
systems and disrupt the orbits of their constituent planets.  This
type of scattering interaction has been studied in the field
\citep{frozen}, and within young embedded clusters (e.g.,
\citealt{adams2006,malmberg2007,malmberg2011,boley2012,dukes};
\citealt{chatterjee,hao2013,pacucci}), where the latter results can be
used to provide constraints on the possible birth environment of our
own solar system (e.g., \citealt{adams2001,hester,willgaidos,spurzem2009}; 
\citealt{zwart,adams2010,williams2010,pfalzner}). We stress that the
dynamical constraints derived for the birth aggegate of the solar
system depend on many variables, including assumptions made about the
cluster properties, any other constraints imposed on the problem, and
the interaction cross sections.

This present study focuses on the cross sections themselves, and
expands previous work to include a much wider range of parameter
space; the implications for the solar birth environment are then
briefly considered at the end of the paper. For studies concerning our
solar system, most previous work has calculated the cross sections for
this mode of disruption by considering the initial orbits of the giant
planets to have their present-day values of semimajor axis.  However,
some recent work suggests that our solar system may have begun in a
more compact configuration \citep{gomes2005,tsiganis}, and the planets
may not have reached their present-day orbits until the solar system
reached an age of hundreds of millions of years. One motivation for
this present study is thus to determine cross sections for solar
system disruption for more compact configurations. Note that the sign
of the effect is not obvious {\it a priori}: The geometrical cross
section of the compact solar system is smaller, and hence implies a
smaller interaction cross section. However, the decreased relative
separations of the planets allow for increased planet-planet
interactions, which could result in more disruption from the passing
stars; in addition, the close spacing in compact solar systems allows
for orbit crossing to occur for smaller values of eccentricity.

In some compact configurations of the solar system, the giant planets
can be at or near mean motion resonance. This possibility leads to
interesting dynamics: Mean motion resonances can protect planetary
systems from disruption, and could thus lead to greater stability and
smaller interaction cross sections. On the other hand, the mean motion
resonances themselves are more easily compromised than planetary
orbits --- the potential energy corresponding to the resonance angle
being in a bound state is much less than the gravitational potential
energy of the planetary orbit. An important related question is thus
to find the cross sections for passing stars (including binaries) to
disrupt mean motion resonances.  Planetary systems with disrupted
resonances will usually retain their planets in the near term,
although they could be subject to orbit instabilities over longer
spans of time.

In addition to compact solar system architectures, this paper
considers a wider range of parameter space than previous studies.
Part of this expanded scope is possible due to increased computational
capabilities. This present study includes results from more than 2
million individual numerical experiments that simulate a solar system
interacting with a passing binary (or single star). For each choice of
solar system architecture and each choice of the background parameters
for the encounters, we run a large ensemble of ${\cal N}_E$
simulations (where ${\cal N}_E$ = 80,000 for most cases, but can be
larger).  The variations that we consider for the target solar systems
include compact configurations (described above), more massive
planets, nonzero initial orbital eccentricities, and a range of masses
for the central stars. Regarding variations in the background
environment, this paper considers two main issues: We determine the
effects of varying the velocity dispersion of the cluster stars, and
we compare the relative sizes of the scattering cross sections for
single stars versus binaries as they interact with planetary systems. 

This paper is organized as follows. We formulate our approach to
calculating the interaction cross sections in Section
\ref{sec:formulate}. The resulting cross sections are then given in
Section \ref{sec:results}, which provides $\cross$ for increases in
eccentricity, increases in the spread of inclination angles, planet
ejection, planet capture, and changes in semimajor axes. Results are
also presented for increasing orbital eccentricities up to
orbit-crossing configurations and compares the efficacy of passing
single and binary stars.  These results are given as a function of
solar system architecture, velocity dispersion of the cluster, and
mass of the host star. Over much of the parameter space of interest,
the cross sections display a nearly self-similar form.  Section
\ref{sec:analysis} presents a scaling analysis that shows how the
results scale with velocity disperion, stellar mass, and starting
semimajor axis. As an application, Section \ref{sec:application}
revisits the possibile dynamical constraints on the birth cluster of
the solar system. In order to assess the level of disruption, one also
needs the rate of close encounters in young stellar clusters. These
rates have already been calculated for a wide range of cluster
properties \citep{adams2006,proszkow2009} and are used herein. The 
paper concludes, in Section \ref{sec:conclude}, with a summary of 
our results and a discussion of their implications. 

\section{Formulation of the Problem} 
\label{sec:formulate} 

One useful way to specify the effects that passing stars can have on
planetary systems is to define cross sections of interaction. For
example, the scattering interactions could eject a planet, increase
the eccentricity, change the semimajor axis, and/or perturb the
inclination angle of the orbit. For a given type of disruption, a
solar system presents an effective target area for being disrupted by
passing stars. With this definition, the effective interaction rate
$\Gamma$ for disruption is then given by the usual formula 
\be
\Gamma = n_\ast \cross \vbar \,,
\label{rate} 
\ee
where $n_\ast$ is the mean density of stars in the environment,
$\vbar$ is the mean relative velocity between systems, and $\cross$ is
the cross section for the given mode of disruption. We note that the
background environment determines the stellar density $n_\ast$ and the
distribution of relative velocities. As outlined below, the relative
velocities follow a Maxwellian distribution characterized by the
expectation value $\vbar$. The interaction cross section depends on
this velocity distribution, so that we actually calculate the quantity
$\cross_v\equiv\langle\sigma{v}\rangle/\vbar$, where the subscript
denotes that the cross section depends on the velocity expectation
values $\vbar$. For ease of notation, however, we drop the subscript
for the remainder of the paper.  In young embedded clusters, we expect
$n_\ast\sim100$ pc$^{-3}$ and $\vbar\sim1-2$ km/s; in the field (in
the solar neighborhood) these quantities have typical values
$n_\ast\sim1$ pc$^{-3}$ and $\vbar\sim30-40$ km/s. Because of the
velocity dependence of the cross sections, solar systems in the field
(with high fly-by speeds) are, on average, less affected by passing
stars.

To calculate the cross sections for interactions, we adopt the
following approach. First we must specify the configuration of the
solar system that will be targetted for disruption (for example, we
can use the current set of four giant planets in our solar system,
with their current masses and semimajor axes, all in orbit about a
solar mass star). Next we must specify the background environment,
which determines the distribution of relative velocities. For most of
this work, we focus on the case where the target solar system
encounters binaries. We then perform a large ensemble of numerical
simulations, where the input parameters are specified according to a
Monte Carlo scheme. The results are then used to calculate the
probability of various outcomes and the corresponding cross sections
(for further detail, see \citealt{frozen,adams2001,adams2006}).

In principle, the Monte Carlo sampling scheme should sample all
possible encounters between binaries and the target solar system,
including those with large impact parameters. In practice, however,
only sufficiently close encounters have a non-negligible chance of
affecting the planetary orbits. In order to conserve computer time, we
thus make the following limitation. We treat the semimajor axes $a$ of
the binaries on a different footing than the others: The values of $a$
are sampled uniformly out to $\amax$ = 1000 AU (more than 30 times the
size of Neptune's orbit in our solar system). For a given value of
$a$, we then limit the possible range of impact parameters to fall
within an area given by $A_0=B\pi a^2$. With this sampling scheme, the
cross section of interaction, for a given type of disruption event,
is given by 
\be
\cross = \int_0^\amax p(a) da \, f_D(a) \, \left(B \pi a^2\right) \,,
\label{crossdef} 
\ee
where $p(a)$ is the probability distribution for binaries having a
semimajor axis $a$. The factor $f_D(a)$ represents the fraction of all
encounters (within the pre-determined area $A_0=B\pi a^2$) that
results in the outcome of interest. Note that the maximum allowed
value of the impact parameter varies with $a$ and is given by
$\varpi_{max}=\sqrt{B}a$. 

The formulation of equation (\ref{crossdef}) can be understood as
follows: Consider a given outcome of interest, say, the ejection of
Neptune. We only consider fly-bys that take place within the area
$A_0=B\pi a^2$, where $a$ is sampled uniformly. If every encounter
within this area leads to the ejection of Neptune, and all encounters
outside this area (which are not computed) have no effect, then
$f_D=1$; the probability factor $p(a)$ corrects for the actual
distribution of binary semimajor axis, and one can see that equation
(\ref{crossdef}) provides the correct effective cross section.  In
practice, of course, only a small fraction of encounters lead to the
ejection of Neptune so that $f_D \ll 1$. As long as we choose the
factor $B$ large enough, we are ignoring only distant encounters that
have little contribution to the cross section.  Nonetheless, since $B$
is finite, this procedure leads to a lower limit on the cross section.
We have run convergence tests with ever-increasing values of $B$ and
find that $B$ = 100 is large enough to include essentially all
relevant encounters. In most of this work we thus use $B$ = 100, which
provides a good compromise between computational speed and accuracy.
For comparison, our previous work \citep{frozen,adams2001} used the
smaller value $B=4$, so that the reported cross sections (again
presented as lower limits) were smaller than those obtained here by a
factor of $\sim2$. This present treatment thus provides a more
complete accounting for wide binaries and results in a greater
lower bound on the true cross sections.

The distribution $p(a)$ is determined by the observed binary period
distribution, which is nearly uniform in the quantity $\log a$, but
has a broad peak centered at period $P = 10^5$ days, which implies
$a \approx 42$ AU for solar type stars \citep{dm}.  

Within the scheme outlined above, encounters between a given solar
system and a passing binary are specified by a large number of input
parameters: We must specify the properties of the binary, including
its semimajor axis $a$, orbital eccentricity $\eb$, the masses of the
two stars $\mone$ and $\mtwo$, and finally the phase of the binary
orbit $\phase$ at the start of the encounter. The orbital elements
$(a,\eb)$ are sampled from their observed distributions \citep{dm}.
Similarly, the stellar masses are sampled from a log-normal form of
the stellar initial mass function (consistent with that advocated by
\citealt{af1996} and \citealt{chabrier}). Both members of the binary
are sampled independently from the distribution and the stellar masses
are limited to the range $M_\ast=0.07-10M_\odot$. As a result, we
exclude brown dwarfs and the very largest stars (which are both rare
and tend to reside at cluster centers).  The phase angle $\phase$ of
the orbit is sampled uniformly over $[0,2\pi]$. Next we must specify
the incoming velocity $v_\infty$ of the solar system with respect to
the binary center of mass; this speed is sampled from a Maxwellian
distribution with a velocity dispersion $\vdisp$ that characterizes
the background environment (e.g., a cluster). The remaining variables
are the three angles $(\theta,\psi,\phi)$ necessary to specify the
direction and orientation of the encounter, and finally the impact
parameter $\varpi$. The impact parameter is chosen randomly within a
circle of radius $10a$ centered on the binary center of mass
(corresponding to the choice $B=100$ in equation [\ref{crossdef}]).

Using a Monte Carlo scheme to select the input parameters according to
the distributions described above, we carry out a large ensemble of
scattering simulations. For most cases we find that the number of
simulations ${\cal N}_E=80,000$ is large enough to provide good
statistics. The outcomes of these numerical experiments are then used
to compute the fraction $f_D$ of disruptive encounters for a given
type of outcome. The resulting errors due to incomplete sampling are
typically 5 percent or less, but can be larger for rare events
(e.g,. for planet ejections, the sampling errors are $\sim10$
percent).

Each simulation is thus an N-body problem. For most cases, $N$ = 7,
where the target system consists of four giant planets orbiting a host
star and interacts with a binary. The equations of motion are
integrated using a Bulirsch-Stoer method \citep{press}, which allows
for rapid integrations and high accuracy.  Because we are interested
in the planetary orbits, which only contain a small fraction of the
total energy of the N-body system, the simulations must conserve the
total energy to high accuracy in order to determine the final orbital
elements. For example, the energy contained in the orbit of Neptune,
the least bound planet, is typically $10^4$ times smaller than the
binding energy of a binary, or the initial gravitational potential
energy between the binary and the solar system. In practice, our
individual simulations have an accumulated error of only one part in
$10^8$, so that orbital changes are safely resolved.

\section{Results for the Cross Sections} 
\label{sec:results} 

Using the formulation described in the previous section, we have
performed several large ensembles of numerical scattering simulations.
Unless stated otherwise, we consider the solar systems to have four
giant planets and to interact with passing binary stars. We then
consider a number of different solar system architectures for the
starting states, as outlined below (see Table 1).  To obtain
reasonable statistics within the Monte Carlo scheme, the number of
individual numerical experiments for each solar system architecture
must typically be of order ${\cal N}_E \approx 80,000$. This choice
produces relative errors (due to incomplete sampling) of order 5
percent or smaller.

\begin{table*}
\begin{minipage}{126mm} 
\label{table1}
\centerline{\bf Solar System Architectures} 
\centerline{$\,$} 
\begin{tabular}{ccccc}
\hline 
\hline
Configuration & Jupiter & Saturn & Uranus & Neptune \\
\hline
Standard & $e=0$ & $e=0$ & $e=0$ & $e=0$ \\ 
Compact  & $a=5.20$ AU & $a=8.67$ AU & $a=14.4$ AU & $a=24.1$ AU \\ 
Resonant & $a=5.88$ AU & $a=7.89$ AU & $a=10.38$ AU & $a=12.01$ AU \\ 
Eccentric \# 1 & $e=0.049$ & $e=0.057$ & $e=0.045$ &  $e=0.011$ \\ 
Eccentric \# 2 & $e=0.10$ & $e=0.10$ & $e=0.10$ &  $e=0.10$ \\ 
Massive & $m_P=1m_J$ & $m_P=1m_J$ & $m_P=1m_J$ & $m_P=1m_J$ \\ 
\hline
\hline
\end{tabular}
\vspace{0.25cm}
\caption{Summary of solar system configurations. In the standard  
configuration (first line), the planets have the same masses and
semimajor axes as those in our solar system but start with zero
eccentricity. For the other configurations, the table entries list the
initial values of the parameters that are different from those of the
standard configuration (so that all of the unlisted parameters have
their standard values).} 
\end{minipage} 
\end{table*}   

\begin{figure} 
\centerline{\psfig{figure=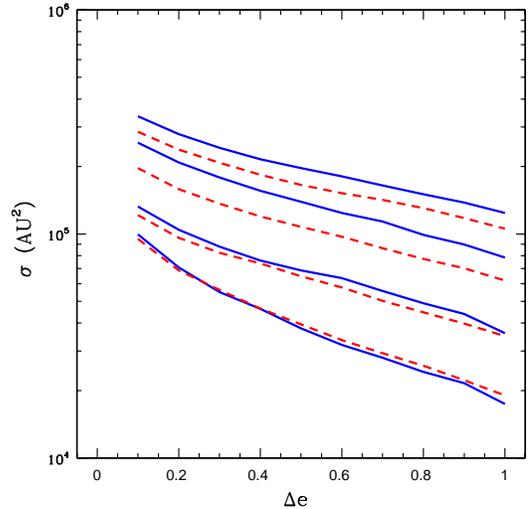,width=8.5cm}} 
\caption{Cross sections for eccentricity increase for the current 
solar system architecture and for a more compact configuration
movitated by the Nice model. For the current solar system (solid blue
curves), the four giant planets are started with their current
semimajor axes and zero eccentricity. For the compact configuration
(dashed red curves), the planets are started with semimajor axes
having a fixed ratio $a_{j+1}/a_j$ = 5/3, where Jupiter ($j=1$) is
started at its present location $a_J$ = 5.2 AU. For both sets of cross
sections, the curves, from top to bottom, correspond to Jupiter
(bottom), Saturn, Uranus, and Neptune (top). Since the orbits start 
with zero eccentricity, the eccentricity increase $\Delta{e}=e$, where
$e$ is the post encounter eccentricity. }
\label{fig:solarcom} 
\end{figure}

In the first set of simulations, we consider the target system to be
an analog of our present-day solar system. In this case, we place the
four giant planets in orbit about a solar mass star and give the
planets their current masses and semimajor axes. The eccentricities
are all set to zero, however, so that we can measure the eccentricity
increases produced by the scattering encounters.  From the results of
these experiments, we compute the cross sections for orbital
disruption of each of the four planets (as outlined in the previous
section).  The results are shown as the solid blue curves in Figure
\ref{fig:solarcom}, which also presents the cross sections for a more
compact starting configuration (described below). The error bars (not
shown) due to incomplete Monte Carlo sampling correspond to relative
errors with a root-mean-square (RMS) value of $\sim$4.4\%.

In Figure \ref{fig:solarcom}, and throughout this paper, the cross
sections for increasing the eccentricity to $e=1$ incorporate all of
the ways that the planet can be removed from its solar system. These
channels include [1] actually increasing the eccentricity to $e\ge1$,
which includes both hyperbolic orbits and planetary orbits that
intersect the host star, [2] ejection from the solar system by
increasing the kinetic energy so that the orbit is unbound, and [3]
capture by one of the (two) passing stars. These channels are not
mutually exclusive, but the simulations are stopped after one of these
events takes place.  However, these channels only include ejection
processes that happen during or immediately after the encounter (we
denote these processes as prompt ejections).  In other cases, the
planets are scattered into high eccentricity orbits, so that the
orbits cross each other. With these configurations, in the absence of
resonance, the planets will eventually experience close encounters,
which in turn lead to ejections or collisions (we denote this process
as delayed ejection).  The cross sections for delayed ejections will
be considered later.

For comparison, we also present the results from a series of numerical
experiments using a more compact orbital architecture (shown as the
red dashed curves in Figure \ref{fig:solarcom}), which is motivated
by the Nice model of solar system formation \citep{gomes2005}.
Although the Nice model has a number of variations, one feature is
that the giant planets could have formed with a more compact
configuration than that of the present day. For this case, we fix the
orbit of Jupiter at $a_J$ = 5.2 AU, and then let each successive
planet have a semimajor axis that is larger than the previous one by a
factor of 5/3. This evenly-spaced solar system thus extends out to
only 24 AU. The results, shown in Figure \ref{fig:solarcom}, indicate
that the cross sections for the compact configuration are somewhat
smaller than those obtained with the current semimajor axes. For this
compact solar system, the RMS errors (not shown) due to incomplete
sampling are $\sim$4.6\%.

\begin{figure} 
\centerline{\psfig{figure=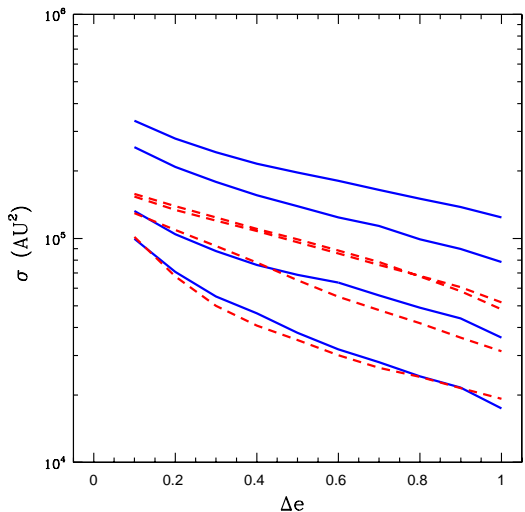,width=8.5cm}} 
\caption{Cross sections for eccentricity increase for the current 
solar system architecture and for a resonant configuration movitated
by the Nice model.  For the current solar system (solid blue curves),
the four giant planets are started with their current semimajor axes
and zero eccentricity. For the resonant configuration (dashed red
curves), the planets are started with semimajor axes $a$ = 5.88, 7.89,
10.38, and 12.01 AU (for the analogs of Jupiter to Neptune). For both
sets of cross sections, the curves, from top to bottom, correspond to
Jupiter (bottom), Saturn, Uranus, and Neptune (top). Since the orbits
start with zero eccentricity, the eccentricity increase $\Delta{e}=e$,
where $e$ is the post encounter eccentricity. }
\label{fig:solarres} 
\end{figure} 

Next we consider an even more compact orbital configuration, again
motivated by the Nice model, where the four giant planets are in
mutual mean motion resonance (MMR). In this case, we choose the
starting semimajor axes to have values of $a$ = 5.88 AU (Jupiter),
7.89 AU (Saturn), 10.38 AU (Uranus), and 12.01 AU (Neptune).  With
these semimajor axes, Jupiter and Saturn are in a 3:2 MMR, Saturn and
Uranus are in a 3:2 MMR, while Uranus and Neptune are in a 5:4 MMR
(for further discussion of this initial state, and others, see
\citealt{batbrown,nesvorny,li2014}). Note that the semimajor axis
ratios do not imply period ratios with exact integer values (although
they are close).  All of the orbital elements must be chosen properly
to put the system in mutual MMR, and this requirement displaces the
period ratios somewhat. Nonetheless, the resonance angles of the
system (for all three planet pairs) are librating in the initial
state, as required for MMR. With this initial state, the solar system
is much more compact than at the present epoch, and the cross sections
for interactions are smaller. This trend is illustrated in Figure
\ref{fig:solarres}, which compares the cross sections with those
obtained for solar systems with the standard starting configuration.
To leading order, the smaller cross sections obtained for the resonant
architecture are a direct consequence of the smaller geometrical size.
However, closer inspection of the results suggests that the cross
sections are larger than the smaller size would imply (see the
analysis of the following section). For example, the cross sections
for changing the eccentricity of Uranus and Neptune are comparable. In
this compact state, planet-planet interactions can be important and
act to increase the the cross sections of Uranus (and Saturn) beyond
the values obtained for more widely separated orbits.

In addition to changes in the orbital elements of the individual
planets, as shown in Figure \ref{fig:solarres}, scattering
interactions can remove solar systems from their resonant states. The
energy required to remove a planetary system from resonance is much
less than that required to eject a planet, or even to substantially
change its orbital elements. To address this issue, we have run an
additional series of numerical simulations to determine the fraction
of systems that are removed from their initial resonant state due to
passing binaries. As before, the ensemble size ${\cal N}_E \approx$
80,000, although the simulations take longer because the resonance
angles must be monitored for several libration times after the
encounters. The result of this set of experiments is the cross section
for removing the solar system from its initial resonant state, namely
\be
\cross_{\rm res} \approx 
\left( 2,280,000 \pm 20,800 \right) {\rm AU}^2 \,.
\label{rescross} 
\ee
This cross section is about 20 times larger than that required to
eject Neptune from the solar system in its normal state, and nearly 40
times larger than the cross section to eject Neptune from the compact,
multi-resonant state. If the removal of the system from resonance
results in orbital instability over longer time intervals, then the
multi-resonant state could be more sensitive to disruption from
passing stars than the standard solar system architecture. We have
carried out 70 longer-term integrations for post-encounter systems and
find that all but one are stable on time scales of $\sim1$ Myr. Other
authors \citep{batbrown,nesvorny} also find that multi-resonant states
can be unstable due to perturbations (generally due to a planetesimal 
disk), and can eject planets, but more follow-up integrations are 
required to assess the probability of significant instability. 

\begin{figure} 
\centerline{\psfig{figure=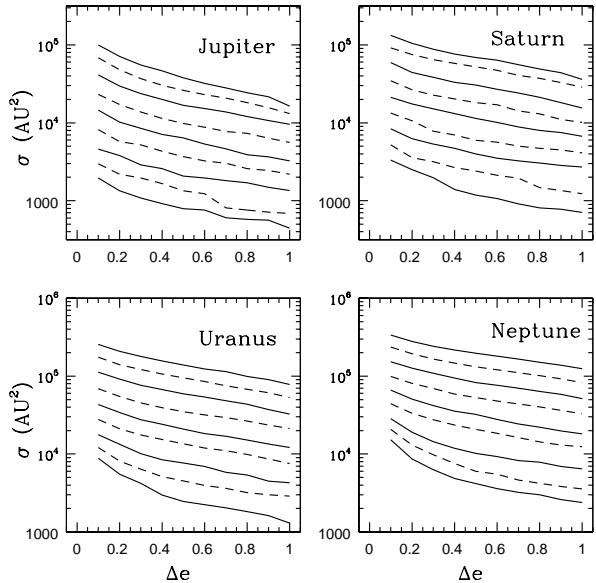,width=8.5cm}} 
\caption{Cross sections for eccentricity increase for the current 
solar system architecture over a wide range of velocity dispersions in
the background cluster. The four giant planets of the solar system
are started with their current semimajor axes and zero eccentricitiy. 
Each panel shows the cross sections to increase orbital eccentricity
for Jupiter (upper left), Saturn (upper right), Uranus (lower left),
and Neptune (lower right). The velocity dispersions fall in the range
from 1 km/s (uppermost curves in each panel) to 16 km/s (lower
curves), and are equally spaced logarithmically (by factors of
$\sqrt{2}$). }
\label{fig:vcross} 
\end{figure} 

The results reported thus far have all been calculated for cases where
the velocity dispersion $\vdisp$ = 1 km/s, a typical value for an
embedded cluster environment \citep{ladalada,porras}. Now we
generalize the treatment by considering the dependence of the cross
section on the velocity dispersion of the background environment. As
is well known, interaction cross sections for high speed encounters,
such as in the field \citep{frozen}, are much lower than those in
clusters \citep{adams2006}, and the velocity dependence is relatively
steep \citep{aspergel,dukes}. To study this dependence, we consider
ensembles of numerical simulations with different values of velocity
dispersion $\vdisp$. More specifically, we consider solar system
starting with the current value of semimajor axes, and $\vdisp$ in the
range from 1 km/s to 32 km/s, varied by factors of $\sqrt{2}$ (so they
are evenly spaced in a logarithmic sense). For the low end of this
range of $\vdisp$, we can use the usual number ${\cal N}_E$ = 80,000
of trials in the ensemble for each value of $\vdisp$. For the larger
values of $\vdisp$, however, the cross sections are lower, and
disruptive events are rare, so that we need larger values of ${\cal
  N}_E$ to obtain good statistics (we find that the choice ${\cal N}_E
\approx 200,000$ is usually large enough).

The interaction cross sections produced by this study are shown in
Figure \ref{fig:vcross}, where each panel corresponds to the results
for one of the giant planets. The cross sections are plotted as a
function of the post-encounter eccentricity $e$, for each choice of
velocity dispersion $\vdisp$. Figure \ref{fig:vcross} shows that the
cross sections are almost evenly spaced in a logarithmic sense, with
the lowest (highest) velocity dispersions producing the largest
(smallest) largest cross sections. This finding suggests that the
cross sections --- to leading order --- display a power-law dependence
on the velocity dispersion. This claim is verified in the following 
section. 

\begin{figure} 
\centerline{\psfig{figure=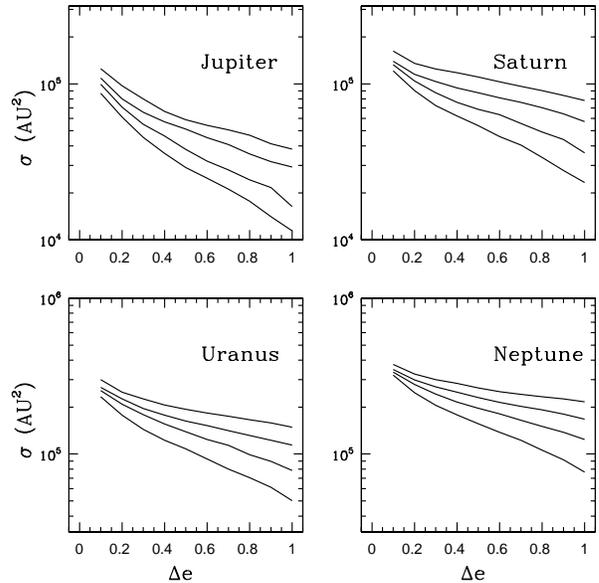,width=8.5cm}} 
\caption{Cross sections for a range of masses of the host star. 
Each case uses analogs of the four giant planets of our solar system,
where the planets start with the current semimajor axes and zero
eccentricitiy.  Each panel shows the cross sections to increase
orbital eccentricity for the analog Jupiter (upper left), Saturn
(upper right), Uranus (lower left), and Neptune (lower right). The
four curves in each panel correspond to four stellar masses, $M_\ast$
= 0.25, 0.5, 1.0, and 2.0 $M_\odot$, from top to bottom. }
\label{fig:mcross} 
\end{figure} 

Next we consider the effect of changing the mass $M_\ast$ of the host
star.  Figure \ref{fig:mcross} shows the cross sections for systems
with the current solar system architecture and varying stellar masses,
from $M_\ast=0.25-2.0M_\odot$. For these numerical experiments, the
solar systems are all started with four planets that have the same
masses and semimajor axes of the giant planets of our solar system.
These analogs are labeled as `Jupiter' through `Neptune', although the
host star can have a mass that differs from the Sun. As expected, the
cross sections shown in Figure \ref{fig:mcross} decrease as the
stellar masses increases. Unlike the case of varying the velocity
dispersion, however, the cross sections, considered as a function of
eccentricity increase, do not display as much self-similarity: The
cross sections decrease more steeply with eccentricity as the mass of
the host star increases. Nonetheless, for a given value of
eccentricity increase, cross sections for the four planets (with their
four values of $a$) all show the nearly same (power-law) scaling with
stellar mass. 

Notice that changing the stellar mass is (in one sense) akin to
changing the planetary masses, because the mass ratios are the most
important variables. However, this association is not an equivalence:
The masses of the passing binaries also enter into the problem, and
their mass distribution is kept invariant. In addition, if the masses
of the planets are increased to the point where the planet-planet
interactions play a role, then self-excitation of eccentricity can
produce larger cross sections. This issue is addressed below where we
consider solar systems with larger planets. We expect interactions to
be important in the regime where the angular momentum exchange time
scale between planets is comparable to the encounter timescale. The
exchange time scale can be determined, but the calculation is
different for widely separated planets where the secular approximation
is valid and for the resonant case (for further discussion, see
\citealt{batmorby}).

\begin{figure} 
\centerline{\psfig{figure=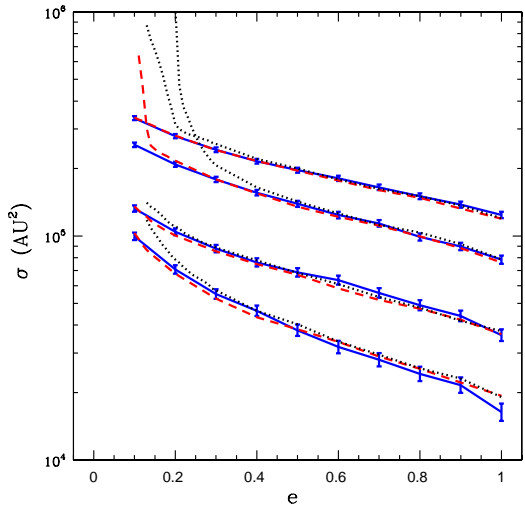,width=8.5cm}} 
\caption{Cross sections for the solar system planets and varying 
initial eccentricities of the planetary orbits. For all cases, the 
four giant planets of our solar system are started with their current
semimajor axes. The solid blue curves show the results for zero
initial eccentricitiy; the dashed red curves show the results where
the planets start with their current orbital eccentricities ($e$ =
0.049, 0.057, 0.045, and 0.011); the black dotted curves show the
results where the starting orbits all have $e$ = 0.10. Cross sections
are given for Jupiter (bottom curves), Saturn, Uranus, and Neptune 
(top curves), all given as a function of the post-encounter value 
$e$ of the eccentricity. } 
\label{fig:efinite} 
\end{figure} 

For the starting configurations used thus far, the initial orbital
eccentricities of the planets have been taken to be zero. Given this
choice, the resulting cross sections represent the cross sections for
increasing eccentricity (which cannot decrease from its initial
value). However, for the related problem of single stars interacting
with binaries, an important difference arises between starting states
where the binary has zero eccentricity and states where the binary has
small but finite eccentricity \citep{heggie}. One might worry that the
cross sections calculated herein could be affected by introducing
small starting eccentricities for the planetary orbits. We have
explored this possibility by using two additional starting
configurations for the solar system. In one case, the planetary orbits
are started with their currently observed eccentricities, $e$ = 0.049,
0.057, 0.045, and 0.011 for Jupiter, Saturn, Uranus, and Neptune,
respectively. In the second case, the planetary orbits are all started
with a larger value of eccentricity $e$ = 0.10. The resulting cross
sections are shown in Figure \ref{fig:efinite}, along with our
previous results with zero starting eccentricity. As shown in the
Figure, all of the cross sections converge to the same values as long
as the final eccentricity is moderately larger than the starting
values.  The difference between results obtained starting with zero
eccentricity and those where the orbits have their current
eccentricity is modest. For the larger starting values $e$ = 0.10,
the cross sections for reaching $e$ = 0.10 are enormous of course,
much larger than the limits of the plot (and hence are not shown).
Even for this starting state, however, the cross sections have almost
converged to their ``natural'' values for $e\gta0.20$, except for the 
case of Uranus; for this planet, the cross sections for eccentricity 
excitation only converge for $e\gta0.35$. 

The results illustrated in Figure \ref{fig:efinite} indicate that the
problem of solar systems interacting with passing binaries is somewhat
different than that of single stars interacting with binaries
\citep{heggie}. Starting with zero eccentricities has a larger effect
in the binary-single-star setting. One difference between the two
cases is that of symmetry: For a single star passing by a binary with
zero eccentricity, the incoming trajectory is the same as the outgoing
trajectory provided that the encounter is distant (so that the binary
orbit can be considered as a ring of mass); this symmetry cancels some
of the forcing. However, this symmetry is absent for solar system
scattering, even when the planetary orbits are circular. The binaries
that impinge upon the solar systems are themselves eccentric, where
$e$ is drawn from the observed binary eccentricity distribution (which
favors high $e$). In addition, the solar systems have four planets,
with different orbital phases, and this property also breaks the
symmetry (albeit to a lesser degree). Another difference between the
two scattering problems is that the cross sections of this paper are
averaged over an ensemble of different binary properties and different
encounter parameters. The binary scattering results \citep{heggie}
show that the the difference between finite eccentricity and circular
orbits is largest for distant encounters, but the effect (the change
in eccentricity) is largest for close encounters (see their Figure
2). The cross sections of this paper include both regimes, but the
cross section is dominated by the close encounters where the results
for $e=0$ and $e\ne0$ are similar.  As a result, starting the
planetary orbits with non-zero eccentricity has only a modest effect
on the cross sections considered in this paper (provided that one
considers post-encounter eccentricities sufficiently larger than the
starting values).

\begin{figure} 
\centerline{\psfig{figure=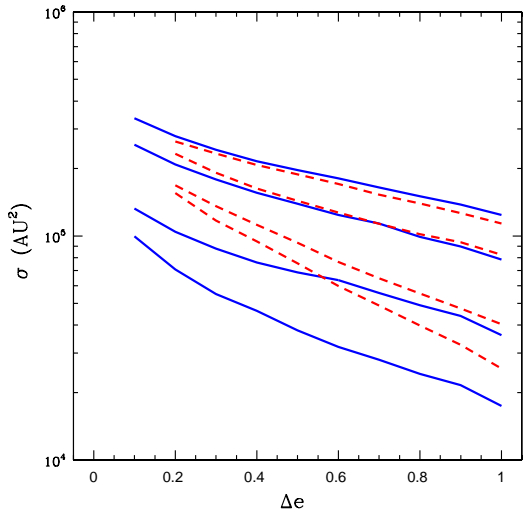,width=8.5cm}} 
\caption{Cross sections for eccentricity increase in systems where 
the giant planets all have mass $m_P=1m_J$ (dashed red curves). The
cross sections for the current solar system architecture are shown for
comparison (solid blue curves).  In both cases, the planets are
started with the current semimajor axes of the giant planets in our
Solar System and with zero eccentricity. Cross sections are shown for
analogs of Jupiter (bottom curves), Saturn, Uranus, and Neptune (top
curves). }
\label{fig:fatcross} 
\end{figure} 

Next we consider the effects of planetary mass on the scattering cross
sections. The results are shown in Figure \ref{fig:fatcross} for the
usual Solar System parameters and for an analog solar system where all
of the giant planets have the mass of Jupiter ($m_P=1m_J$). Both
classes of systems start with the same semimajor axes (the present-day
values in our system) and zero eccentricity.  The figure shows that
the cross sections for increasing the eccentricities of Neptune and
Uranus are largely unaffected by the increase in planetary mass, but
the cross sections for Jupiter and Saturn are somewhat larger.  Note
that the cross sections are plotted only for eccentricity values
$e\ge0.20$. Within such a massive planetary system, small
eccentricities ($e\sim0.10$) are easily excited by planet-planet
interactions; as a result, the cross sections for eccentricity
increase --- as determined through our numerical scheme --- are
extremely large and are not plotted in the figure. 

The numerical results for the cross sections can be understood as
follows: To leading order, we often expect the planets to act as test
particles, so that the cross sections should not be sensitive to the
planetary masses. For sufficiently massive planets, however, an
increase in the eccentricity of one planet can lead to significant
perturbations acting on the other planets, thereby leading to
increased eccentricity excitation. By increasing the mass of all of
the planets to that of Jupiter, the resulting solar systems are more
excitable. The largest increase in the cross sections, which occurs
for Jupiter and for low eccentricties, is only a factor of $\sim2$;
most cross sections experience smaller changes.  These results are
generally consistent with the idea that our Solar System is ``full'',
i.e., no additional planets and little additional mass can be added to
the extant planets without rendering the system unstable. In fact,
even the current solar system is unstable on sufficiently long time
scales \citep{batlaugh,laskar}. 

\begin{figure} 
\centerline{\psfig{figure=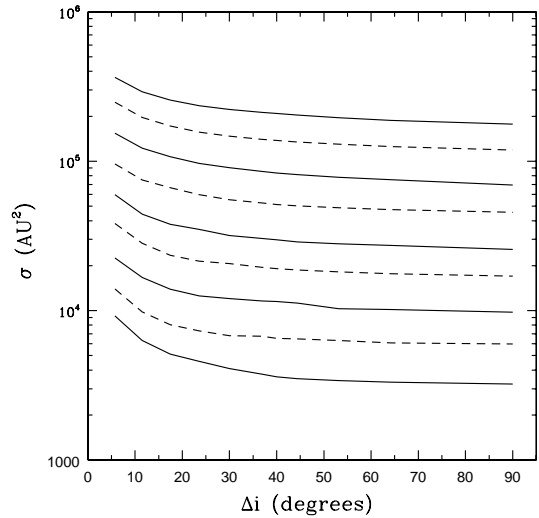,width=8.5cm}} 
\caption{Cross sections for increasing the spread of inclination 
angles of the planetary orbits. All of the giant planets are started
in the same plane; the quantity $\Delta i$ is the total range of
inclination angles of the four orbits after the encounters. Cross
sections are shown for a variety of velocity dispersions, from
$\vdisp$ = 1 km/s (top curve) to $\vdisp$ = 16 km/s (bottom curve),
where the values are evenly spaced logarithmically (by factors of
$\sqrt{2}$). }  
\label{fig:icross} 
\end{figure} 

Another way in which planetary orbits can be altered by scattering
encounters is by changing their inclination angles. For all of the
simulations, we start the four giant planets in the same plane (so
that $i_J=i_S=i_U=i_N=0$). After the encounters, the inclination
angles of the four planets are, in general, nonzero. We define the
post-encounter spread $\Delta i$ of the inclination angles according 
to the expression 
\be
\Delta i \equiv {\rm max} \left\{ \cos^{-1} \left[ 
{ {\bf J}_j \cdot {\bf J}_k \over J_j J_k} \right] \right\} \,,
\ee
where the ${\bf J}_i$ are the angular momentum vectors of the
planetary orbits and where the indices run through all four of the
giant planets. The resulting cross sections for increasing the spread
of inclination angles is shown in Figure \ref{fig:icross}. The Figure
shows the cross sections for a range of velocity dispersions of the
background cluster, from $\vdisp$ = 1 km/s to $\vdisp$ = 16 km/s,
where the values are spaced by factors of $\sqrt{2}$. The cross
sections are almost evenly spaced in the semi-logarithmic plot and
have nearly the same shape as a function of $\Delta i$.  These
properties indicate that the cross section has a power-law dependence
on $\vdisp$ (see Section \ref{sec:analysis}).

\begin{figure} 
\centerline{\psfig{figure=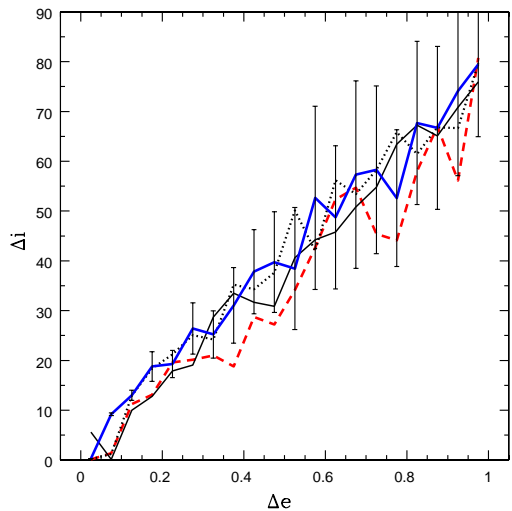,width=8.5cm}} 
\caption{Correlation between eccentricity increases and increases 
in the spread of inclination angles of the planetary orbits.  All of
the giant planets are started in the same plane with circular orbits;
the quantity $\Delta i$ is the total range of inclination angles of
the four orbits after the encounters. Correlations are shown for the 
orbital elements changes of Jupiter (heavy dashed red curve), Saturn
(black solid curve), Uranus (black dotted curve), and Neptune (heavy
blue solid curve). For each planet, the inclination angle increases
are binned over a range in $\Delta e$ of width $\delta=0.05$. Although
the correlation is well-defined, the range of $\Delta i$ for a given
value of $\Delta{e}$ is relatively large. The error bars (shown for
the Neptune curve only) depict the standard deviations. }
\label{fig:ive} 
\end{figure} 

In general, increases in the inclination angles are positively
correlated with increases in eccentricity. This result is not
unexpected, as changes in both orbital elements correspond to
disruption of the initial states. To illustrate this trend, in Figure
\ref{fig:ive} we plot the increases in the spread of inclination angle
$\Delta i$ versus the change in eccentricity (equivalently, the
post-encounter eccentricity since $\Delta{e}=e$). The two variables
are in fact well correlated, but the range of possible $\Delta i$
values for a given eccentricity $e=\Delta{e}$ is large. As a result,
in the figure we plot the mean values of $\Delta i$ averaged over a
bin in $\Delta e$ with a width of $\delta$ = 0.05. The data show a
well-defined correlation; for this choice of binning, the spread in
the inclination angles grows to about 80$^\circ$ as the eccentricity
grows to unity.  The four curves shown in Figure \ref{fig:ive}
correspond to the four giant planets. Note that the orbits of all four
planets show the same general trend.

\begin{figure} 
\centerline{\psfig{figure=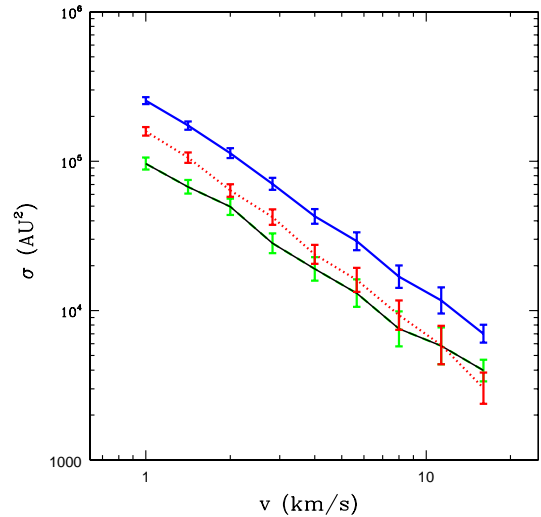,width=8.5cm}} 
\caption{Cross sections for the ejection of at least one planet 
as a function of velocity dispersion $\vdisp$ in the cluster. The
target systems have four giant planets with the masses and semimajor
axes of our solar system bodies. Cross sections are shown for three
cases: increases in eccentricity large enough to produce orbit
crossing (solid dark curve marked by green error bars), direct
ejection of a planet (dotted red curve), either channel of ejection
(solid blue curve). }
\label{fig:eject} 
\end{figure} 

The cross sections discussed thus far correspond to the immediate,
post-encounter properties of the solar systems. In addition to
immediate ejection, however, the solar systems can be rendered
sufficiently unstable so that they eject planets long after the
scattering encounters are over. These longer term ejection events can
be divided into (at least) two types. In the first --- and most
unstable --- case, the scattering encounters leave the planetary
orbits with high enough eccentricity so that adjacent orbits cross.
Most orbiting-crossing systems will eventually eject one of their
planets, provided that the system is not in a mean motion resonance;
furthermore, perturbations due to stellar encounters are unlikely to
place a planetary system in resonance. We address the effects of this
type of instability by finding the cross sections for producing
orbit-crossing planetary systems (see below). In the second case,
systems with more modest eccentricities can be unstable over long
spans of time. In order to assess the effects of this latter class of
outcomes, the post-encounter systems must be integrated over typical
stellar ages (billions of years). This task is beyond the scope of
this present work, but provides an interesting problem for the future.

Using the results of our numerical experiments, we can calculate the
cross sections for the scattering interactions to leave any two orbits
with high enough eccentricities to cross. For the case of the analog
solar system, where the four giant planets have their current masses
and semimajor axes, the resulting cross sections are shown in Figure
\ref{fig:eject}. Three sets of cross sections are shown as a function
of the velocity dispersion $\vdisp$ of the background cluster. The
cross sections for the post-encounter system to have an orbit-crossing
configuration are shown as the lower, green solid curve in the figure.
For the calculation of this cross section, only systems where all of
the planets are retained by the host star are included. The cross
sections for the system to eject any planet (including those planets
captured by the passing stars) are shown as the red dotted curve.
Finally, the total cross sections for ejection, including both direct
ejection of a planet and/or crossing orbits, are shown as the blue
solid curve in the figure.  The error bars depict the uncertainties in
the cross sections due to incomplete Monte Carlo sampling. Note that
the cross sections for orbit crossing and the cross sections for
direct ejection are roughly comparable, with the latter slightly
larger (except at high velocity dispersion, where they are the same
within the sampling uncertainties).  The total cross section for
ejection is thus larger than that for direct ejection by a factor of
$\sim2$. This statement holds only for the current solar system
architecture, but remains valid over the range of velocity dispersion
shown here ($\vdisp=1-16$ km/s).

We can now compare the results for the standard solar system
architecture with that of the more compact configuration motivated by
the Nice model. Here we consider only the most compact version where
the planets are in multiple mean motion resonances (see Figure
\ref{fig:solarres}). The compact configuration is expected to have
lower cross sections for direct ejection. But the orbits are closer
together, so that less eccentricity excitation is required to produce
crossing orbits.  On the other hand, the semimajor axes are smaller,
which lowers the cross sections for eccentricity increase. We find
here that the cross sections for orbit crossing are comparable,
$\cross = 96,500 \pm 3750$ AU$^2$ for the standard configuration
versus $\cross = 92,200 \pm 3710$ AU$^2$ for the compact
multi-resonant case. However, the cross section for direct ejection is
larger for the standard solar system by a factor of 1.5, so that the
total ejection cross section remains larger by a factor of $\sim1.25$.

\begin{figure} 
\centerline{\psfig{figure=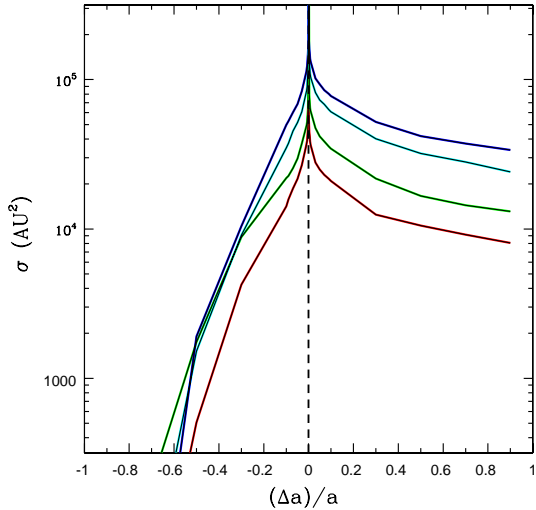,width=8.5cm}} 
\caption{Cross sections for changes in the semimajor axis of the  
planetary orbits due to scattering encounters. The target systems are
analogs of our Solar System, with the four giant planets initially in
circular orbits with the current values of their semimajor axes. The
plots shows the cross sections for relative changes $(\Delta a)/a$ in
the semimajor axis for the orbits of Jupiter (lower red curve), Saturn
(green curve), Uranus (cyan curve), and Neptune (upper blue curve). } 
\label{fig:across} 
\end{figure} 

Although the semimajor axes of planetary orbits are altered less
dramatically than the eccentricities and inclination angles during
scattering encounters, the values of $a$ are nonetheless affected.
The possible variations are quantified in Figure \ref{fig:across},
which shows the cross sections for producing relative changes
$(\Delta{a})/a$ in the semimajor axes of the four giant planets.  
This ensemble of numerical simulations uses the standard solar system
architecture as initial conditions, where the planets have their
current masses and semimajor axes. The velocity dispersion of the
background cluster is taken to be $\vdisp$ = 1 km/s. As expected, the
cross sections are largest for Neptune (top blue curve) and smallest
for Jupiter (bottom red curve). As a crude approximation, the cross
sections are proportional to the starting semimajor axes of the
planets (although closer inspection shows the scaling is somewhat less
steep than linear).

Scattering encounters can cause the semimajor axes to become either
smaller or larger, corresponding to the loss or gain of orbital
energy. However, Figure \ref{fig:across} shows that the process is
highly asymmetric, where the orbits are much more likely to become
larger (gain energy) than to move inward (lose energy). The scattering
encounters rarely reduce the semimajor axes by more than a factor of
two.  Moreover, the magnitude of the cross sections are relatively
small. More specifically, the cross sections for changing the initial
semimajor axes by 10\% are roughly comparable to --- but somewhat
smaller than --- the cross sections for ejecting a planet (compare
Figures \ref{fig:solarcom} and \ref{fig:across}). One might think that
cross sections for moderate changes $\Delta a$ would be larger than
those for ejection. However, the cross sections for changes in
semimajor axis do not include the ejections themselves, i.e., they are
the cross sections for changing the semimajor axis with the planet
remaining bound to its host star. For large changes in $a$, there is
not much parameter space where $a$ is increased but the planet remains
bound (thereby leading to the values shown in Figure \ref{fig:across}).  
Notice also that the figure does not show cross sections for overly
small values of $(\Delta a)/a$; the cross sections become singular in
the limit $(\Delta a)/a \to 0$, as marked by the vertical dashed line.

\begin{figure} 
\centerline{\psfig{figure=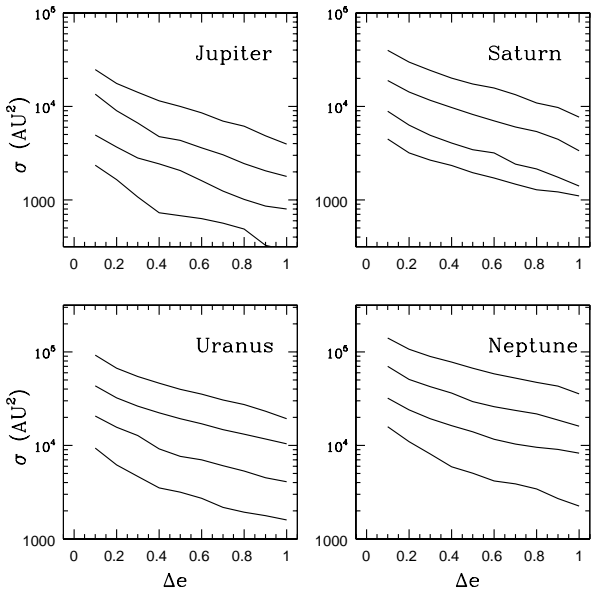,width=8.5cm}} 
\caption{Cross sections for eccentricity increase due to encounters  
with passing single stars. The target systems are analogs of our Solar
System, with the four giant planets in circular orbits with their
current values of semimajor axis. Each panel shows the cross sections
for a given planet, as labeled, where the curves correspond to varying
velocity dispersions of the background cluster: $\vdisp$ = 1, 2, 4,
and 8 km/s (from top to bottom). }
\label{fig:singlecross} 
\end{figure}

The cross sections considered thus far correspond to interactions
between solar systems and passing binaries. On the other hand, roughly
half of the stellar population consists of single stars, so that the
corresponding cross sections for singles must also be determined.
Since we are primarily interested in a comparison between the cross
sections for single stars and binaries, it is crucial to use the same
sampling for all of the parameters in the problem.  Toward this end,
we use exactly the same procedure as before (outlined in Section
\ref{sec:formulate}), but let the mass of the second star go to
zero. In this limit, the other, single star automatically resides at
the center of mass of the system (and the value of the binary
eccentricity becomes irrelevant).  The resulting cross sections for
single stars interacting with solar system analogs are shown in Figure
\ref{fig:singlecross}. As before, the initial solar systems consist of
four giant planets with the masses and semimajor axes of the present
day Solar System (but with zero starting eccentricity). Each panel
shows the interaction cross sections for eccentricity increases for a
given planet (as labeled). Results are shown for four values of the
velocity dispersion of the background cluster, i.e., $\vdisp$ = 1, 2,
4, and 8 km/s (ordered from top to bottom in each panel).

Next we make a rough comparison of the cross sections for single star
interactions (Figure \ref{fig:singlecross}) with those obtained
earlier for binaries (e.g., Figure \ref{fig:vcross}). The single star
cross sections are smaller by {\it more than a factor of two}. Note
that the binary systems are, on average, somewhat wider than the size
of the solar systems. As a result, as a pair of stars passes by a
solar system, it consists mostly of empty space but still provides
(roughly) twice the opportunity for interaction as a single star. One
thus expects at least a factor of two reduction in the cross sections
for passing singles. The fact that the reduction is larger than a
factor of two is thus significant and indicates that the dynamics of
the binaries themselves must contribute. Further, as discussed in the
following section, the cross sections for single stars exhibit a
different dependence on the background velocity dispersion and a
slightly steeper dependence on post-encounter eccentricity. 

For convenience, Table 2 collects the cross sections for the ejection
and capture of all four planets.  For each solar system configuration
considered in this paper, the table lists two sets of cross sections,
where the first line corresponds to planetary ejection and the second
line corresponds to planetary capture. The Standard Model (the first
configuration in the table) represents the case where the four giant
planets have the masses and semimajor axes of our current Solar
System, the host star has mass $M_\ast=1.0M_\odot$, the velocity
dispersion of the cluster $\vdisp=1$ km/s, and the interacting stars
are binary. The first column in the table labels the solar system
configuration by the variable that differs from its standard value.
The error bars in the table are those due to incomplete Monte Carlo
sampling. One way to assess statistical significance is through the
ratio of the cross section to its sampling error.  For the ejection
cross sections, the mean value (averaged over the entire table) of
this signal to noise ratio is $\sim14$, so that the ejection cross
sections are well-determined. Capture events are much more rare. For
the capture cross sections, the mean value of the signal to noise
ratio is only $\sim4$. For the rarest events, captures with high
cluster velocity dispersion, the cross sections are only defined at
the factor of two level. 

\begin{table*}
\begin{minipage}{126mm} 
\label{table2}
\centerline{\bf Cross Sections for Ejection and Capture}
\centerline{$\,$} 
\begin{tabular}{ccccc}
\hline 
\hline
Configuration & Jupiter & Saturn & Uranus & Neptune \\
\hline
Standard model & 
15500 $\pm$ 1360 & 34000 $\pm$ 2050 & 72300 $\pm$ 3100 & 113000 $\pm$ 3860 \\
& 812 $\pm$  306 &  2140 $\pm$  531 &  6040 $\pm$  994 & 11400 $\pm$ 1280 \\
Compact model & 
18100 $\pm$ 1510 & 32700 $\pm$ 2130 & 57500 $\pm$ 2790 & 93900 $\pm$ 3570 \\
& 915 $\pm$  379 &  2280 $\pm$  607 &  4380 $\pm$  817 & 11500 $\pm$ 1320 \\
Resonant model & 
23900 $\pm$ 1810 & 40200 $\pm$ 2440 & 61100 $\pm$ 2990 & 60100 $\pm$ 2900 \\
&1240 $\pm$ 467  &  2150 $\pm$  569 &  3430 $\pm$  701 &  3620 $\pm$  738 \\ 
Massive planets &
24100 $\pm$ 1890 & 38300 $\pm$ 2390 & 77700 $\pm$ 3360 & 105000 $\pm$ 3880 \\
&1530 $\pm$  579 &  2170 $\pm$  637 &  4810 $\pm$  932 &  8480 $\pm$ 1190 \\
\hline  
$\vdisp$ = 2 km/s & 
9170 $\pm$ 947 & 14800 $\pm$ 1250 & 29800 $\pm$ 1770 & 45200 $\pm$ 2240 \\
&391 $\pm$ 136 &   635 $\pm$  173 &  2600 $\pm$  487 &  6370 $\pm$  903 \\
$\vdisp$ = 4 km/s & 
2980 $\pm$  454 &  6090 $\pm$  776 & 10600 $\pm$  918 & 15700 $\pm$ 1140 \\
&270 $\pm$  173 &   607 $\pm$  230 &  1580 $\pm$  486 &  2430 $\pm$  569 \\
$\vdisp$ = 8 km/s & 
1220 $\pm$  258 &  2580 $\pm$  403 &  4060 $\pm$  506 &  5830 $\pm$  698 \\
&130 $\pm$   85 &   134 $\pm$   75 &   239 $\pm$   88 &   624 $\pm$  228 \\
$\vdisp$ = 16 km/s & 
181 $\pm$  39 &   607 $\pm$ 113 &  1220 $\pm$ 182 &  2140 $\pm$ 252 \\
&82 $\pm$  52 &    53 $\pm$  39 &   214 $\pm$  83 &   169 $\pm$  69 \\ 
\hline 
$M_\ast=0.25M_\odot$ &  
37400 $\pm$ 2195 & 74800 $\pm$ 3170 & 138000 $\pm$ 4350 & 196000 $\pm$ 5130 \\
& 766 $\pm$  330 &  3510 $\pm$  742 & 10300 $\pm$ 1240 & 19500 $\pm$ 1720 \\
$M_\ast=0.5M_\odot$ &  
27600 $\pm$ 1830 & 54000 $\pm$ 2630 & 107000 $\pm$ 3740 & 152000 $\pm$ 4460 \\
&1730 $\pm$  560 &  3310 $\pm$  755 &  7470 $\pm$ 1060 & 15100 $\pm$ 1490 \\
$M_\ast=2.0M_\odot$ &  
11000 $\pm$ 1170 & 21700 $\pm$ 1720 & 45700 $\pm$ 2420 & 69300 $\pm$ 2980 \\
& 458 $\pm$  234 &  1590 $\pm$  467 &  4420 $\pm$  844 &  7410 $\pm$ 1110 \\
\hline 
Single, $\vdisp=1$ km/s & 
3840 $\pm$  651 &  7100 $\pm$  856 & 17300 $\pm$ 1430 & 30600 $\pm$ 1980 \\
&135 $\pm$   80 &   587 $\pm$  210 &  2080 $\pm$  480 &  5090 $\pm$  871 \\
Single, $\vdisp=2$ km/s & 
1620 $\pm$  324 &  3110 $\pm$  429 &  9030 $\pm$  926 & 13200 $\pm$ 1090 \\
&168 $\pm$   86 &   236 $\pm$  106 &  1370 $\pm$  423 &  2810 $\pm$  641 \\
Single, $\vdisp=4$ km/s & 
 685 $\pm$  177 &  1300 $\pm$  244 &  3740 $\pm$  531 &  6790 $\pm$  793 \\
&116 $\pm$   94 &   117 $\pm$   51 &   360 $\pm$  117 &  1480 $\pm$  411 \\
Single, $\vdisp=8$ km/s & 
 269 $\pm$  103 &  1090 $\pm$  322 &  1440 $\pm$  286 &  1880 $\pm$  266 \\
& 21 $\pm$   14 &    23 $\pm$   14 &   157 $\pm$   53 &   374 $\pm$  202 \\
\hline
\hline
\end{tabular} 
\vspace{0.25cm}
\caption{For each solar system configuration, as labeled in the left 
column, the top line lists the ejection cross sections and the second
line lists the capture cross sections for each of the planets. The
error bars due to incomplete Monte Carlo sampling are included. All
cross sections are given in units of AU$^2$. }
\end{minipage} 
\end{table*}

\section{Analysis and Scaling Laws} 
\label{sec:analysis} 

The cross sections found in the previous section display relatively
simple dependences on the underlying variables of the problem: For
example, for each type of solar system, the cross sections, when
considered as functions of the post-encounter planetary eccentricity
$e$, all display the same general shape. As a result, the functions
$\cross(e)$ can (almost) be rescaled to find a universal functional
form, where scaling factors take into account the initial semimajor
axis $a$ of the planet, the velocity dispersion $\vdisp$ of the
background environment, the stellar mass $M_\ast$, and so on. The goal
of this section is to understand the general scaling properties of the
cross sections and to determine the extent to which they are
self-similar. In general, self-similarity arises when physical scales
are either missing from a problem or do not contribute to the results
\citep{barenblatt}; we return to this issue at the end of the section.

Even in the reduced case where we consider one planet at at time, the
interactions considered in this paper involve four bodies (the host
star, the planet, and two binary members). Unfortunately, four-body
interactions are rather difficult to describe analytically to any
reasonable degree of approximation. As a result, the goal of this
section is relatively modest: Instead of building complicated
analytical models for 4-body (and higher N-body) dynamics, we consider
here basic physical principles that can be used as motivation for
scaling laws. We then combine these heuristic results with our
detailed numerical determinations of the cross sections. The result 
is physically motivated fitting formula that characterize the cross
sections over the parameter space of interest $(a,\vdisp,M_\ast,e)$.

To start the discussion, consider the simplest case where the the
cross section for interactions is the geometrical cross section
$\pi{a^2}$ provided by a planet in its initial orbit. Further, we
consider the planets to be independent of each other during the
encounters.  This cross section will be enhanced by gravitational
focusing, so we can write down an heuristic expression for the cross
section in the form 
\be
\cross_0 \approx \alpha \pi a^2  
\left( 1 + {v_{esc}^2 \over v_\infty^2} \right) \,,
\label{crossguess} 
\ee
where $v_{esc}$ is the escape speed from the target system (at the
location of the planet), and $v_\infty$ is the asymptotic relative
speed between the two systems. In order to pass within this cross 
sectional area, the interacting star (binary) must be about the same  
distance from the planet as its host star, so that the planet has a
chance of being ejected from its original solar system. This
expression thus represents the escape cross section. The parameter
$\alpha$ is a dimensionless constant of order unity and is included 
to encapsulate the uncertainties inherent in this approximation. 
After inserting the expression for the escape speed, we obtain
\be
\cross_0 = \alpha \pi a^2 
\left( 1 + {G M_\ast \over a \vdisp^2} \right) \to 
\alpha \pi \ell a \,, 
\label{crosslinear} 
\ee
where we have replaced the asymptotic speed $v_\infty$ with the
velocity dispersion $\vdisp$ of the cluster (or other background
stellar system) and we have defined the corresponding length scale
$\ell \equiv GM_\ast/\vdisp^2$ (where $\ell\sim890$ AU for $\vdisp=1$
km/s). The final expression represents the limiting form, which is
applicable when gravitational focusing dominates, and implies a linear
dependence of the cross section on $a$. Given this form, the cross
section requires another length scale. In this problem, the orbit
speed of the binary, the asymptotic speed $v_\infty$ of the encounter,
the orbit speed of the planet, and the velocity dispersion $\vdisp$
are all roughly comparable (1 -- 10 km/s). For example, the orbit of
Neptune in our solar system has an escape speed of $\sim5.5$ km/s,
whereas the orbit speed of a binary at the peak of the period
distribution is also $\sim5$ km/s. If we only have a single velocity
$V$, then dimensional analysis implies that the relevant length scale
must be $\ell = GM_\ast/V^2$, as given in equation 
(\ref{crosslinear}); additional uncertainties can be absorbed into the
dimensionless parameter $\alpha$. Finally we note that for a velocity
dispersion $\vdisp$ = 1 km/s, the gravitational focusing term
dominates by a factor of 30.

\begin{figure}  
\centerline{\psfig{figure=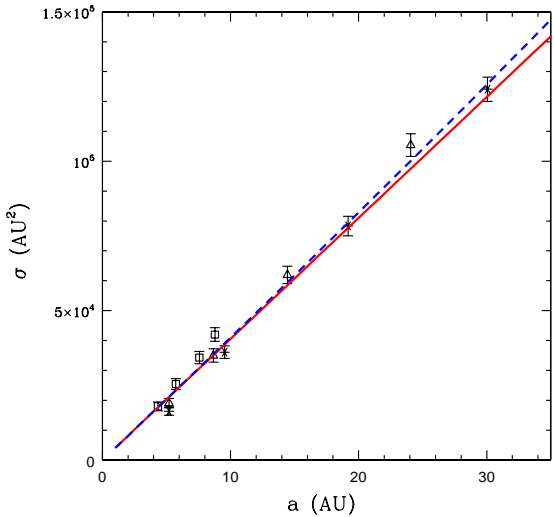,width=8.5cm}} 
\caption{Cross sections for planetary escape versus the starting 
semimajor axis. The 12 points on the plot correspond to the four giant
planets in each of three versions of the initial solar system
architecture. The symbols represent different starting states,
including the semimajor axes of the present-day solar system (stars),
a compact configuration with 5/3 semimajor axis ratios (open
triangles), and an ultra-compact solar system starting in multiple
mean motion resonances (open squares).  The red solid line shows the
cross section indicated by the limiting form of equation
(\ref{crosslinear}); the blue dashed curve shows the full form. } 
\label{fig:clinear} 
\end{figure}

The limiting form of equation (\ref{crosslinear}) is linear in the
starting semimajor axis $a$ of the planet. To see how well this
expression works, we plot the ejection cross sections of the planets
versus semimajor axis in Figure \ref{fig:clinear}. As expected, the
ejection cross section is a nearly linear function of the semimajor
axis. This trend holds for solar systems starting with the present-day
semimajor axes (star symbols) and the more compact configuration where
the semimajor axes are spaced by factors of 5/3 (open triangles). We
also plot results for ultra-compact solar systems in multiple mean
motion resonance (open squares). In order to isolate the dependence of
the cross sections on initial semimajor axis from planet-planet
scattering effects, this latter case uses smaller planet masses (by a
factor of 10), so they act more like test masses; to explore a wider
range in $a$, we also take this compact system to be smaller by a
factor of 1.35 compared to that considered in the previous section.
The error bars delineate the uncertainty due to incomplete Monte Carlo
sampling. Not only do the cross sections show nearly linear dependence
on $a$, but the slope of the curve is predicted by the above analysis.
The red solid (blue dashed) curve in Figure \ref{fig:clinear} shows
the cross section predicted by equation (\ref{crosslinear}) for the
limiting case (full form); for both cases, the characteristic length
scale $\ell$ = 890 AU and the dimensionless parameter $\alpha =
7/5$.\footnote{In order to set the value for this dimensionless
  parameter, and others specified in this section, we generally search
  in increments of $10^{-2}$, find the value that gives the minimum
  RMS error, and then choose the nearest round number (ratio of
  relatively small integers). }

\begin{figure}  
\centerline{\psfig{figure=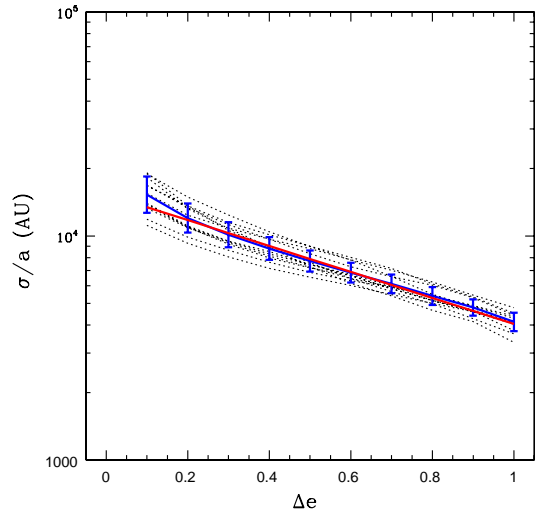,width=8.5cm}} 
\caption{Scaled cross sections versus eccentricity increase $\Delta{e}$
(equivalently, the post-encounter eccentricity $e$) for the four giant  
planets in each of the starting architectures for the solar system. 
The individual cases are shown as light dotted curves. The heavy solid
blue curve depicts the average, where the error bars depict the
standard deviation. The straight red line shows the result for cross
sections with a purely exponential dependence on eccentricity. }
\label{fig:scale} 
\end{figure}

Next we consider the dependence of the cross sections on the
post-encounter eccentricity $e$ (which is equivalent to $\Delta{e}$
because the orbits start with zero eccentricity). For all four planets
in all three types of solar system, the $e$-dependence is similar.
Since the ejection cross sections scale linearly with semimajor axis
$a$ (see Figure \ref{fig:clinear}), we scale the cross sections by
dividing out one power of $a$. The resulting scaled cross sections are
shown in Figure \ref{fig:scale} as a function of eccentricity $e$. In
addition to the individual cases (shown as the light dotted curves),
the average is shown as the heavy blue curve, where the error bars
depict the standard deviation. This latter quantity provides a measure
of the spread in the values of the cross section over the various
cases. The standard deviation varies from about 17\% of the cross
section at low eccentricity $e$ = 0.10 to only about 9\% at $e$ = 1.0.

The curves in Figure \ref{fig:scale} are nearly straight lines on the
semi-logarithmic plot, so that the dependence of the cross sections on
eccentricity is nearly exponential. For purposes of illustration, we
use an exponential fitting function of the form 
\be
{\cross_e \over a} = \alpha \pi \ell \exp [\,b\,(1-e)\,]\,, 
\label{expeccent} 
\ee
where the first factor enforces consistency with the ejection cross
sections considered above.  For the value $b$ = 4/3, we obtain a good
fit to the calculated, scaled cross sections, as shown by the heavy
red line in Figure \ref{fig:scale}. Except for first point ($e$ =
0.1), the exponential fit (straight red line) agrees with the average
values (solid blue curve) to within about 3\%, i.e., the difference is
much less than the width of the distributions as measured by the
standard deviations. Another measure of the quality of the fit is
provided the relative differences between the numerically determined
cross sections used in constructing Figure \ref{fig:scale} and the
exponential form given by equation (\ref{expeccent}); the RMS of these
relative errors is $\sim12\%$.

\begin{figure}  
\centerline{\psfig{figure=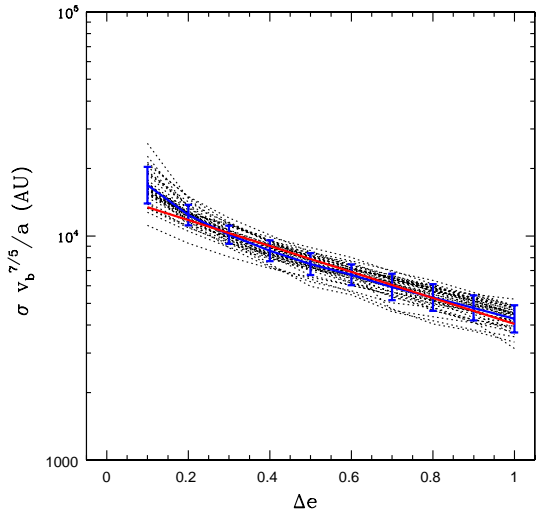,width=8.5cm}} 
\caption{Scaled cross sections versus eccentricity increase $\Delta{e}$
(equivalently, post-encounter eccentricity $e$) for a collection of 
different velocity dispersions for the background. The starting state
is taken to have four giant planets with the current semimajor axes.
Cross sections are scaled by $\vdisp^{7/5}/a$ (see text).  The
individual cases are shown as light dotted curves, with include curves
for each of the planets for $\vdisp$ = 1 -- 16 km/s, equally spaced
logarithmically (by factors of $\sqrt{2}$).  The heavy solid blue
curve depicts the average, where the error bars depict the standard
deviation. The red striaght line shows the result for cross sections
with a purely exponential dependence on eccentricity. }
\label{fig:vscale} 
\end{figure}

Next we consider the effects of the velocity dispersion of the 
background cluster environment. As shown in Figure \ref{fig:vcross} in
the previous section, the cross sections vary with the post-encounter
eccentricity with approximately the same functional form over a wide
range of $\vdisp$. Only the leading cofficient changes. Moreover, the
uniform spacing of the curves in Figure \ref{fig:vcross} indicates
that the cross sections must have a power-law dependence on the
velocity dispersion $\vdisp$ (to leading order).  We have explored
scalings with velocity dependence of the form $\cross \propto
\vdisp^{-\gamma}$ and find that the best fit occurs for $\gamma
\approx 7/5$. Using this choice of power-law index, we plot the scaled
cross sections versus post-encounter eccentricity in Figure
\ref{fig:vscale}, where we include the linear $a$-dependence found
previously (i.e., $\cross \vdisp^{7/5}/a$). Each light dotted curve in
the figure shows the result for one planet and one choice of velocity
dispersion. The heavy blue curve shows the average over all of the
curves, where the error bars depict one standard deviation. The heavy
straight red line represents the same exponential dependence given in
equation (\ref{expeccent}) and used in Figure \ref{fig:scale}. The RMS
of the relative differences between the numerically determined cross
sections and the curve given by equation (\ref{expeccent}) is
$\sim13\%$. The cross section curves are thus self-similar to this
level of accuracy. Furthermore, the dependence of the cross sections
on velocity dispersion is nearly independent of the dependence on
starting semimajor axis $a$ of the planet.

\begin{figure} 
\centerline{\psfig{figure=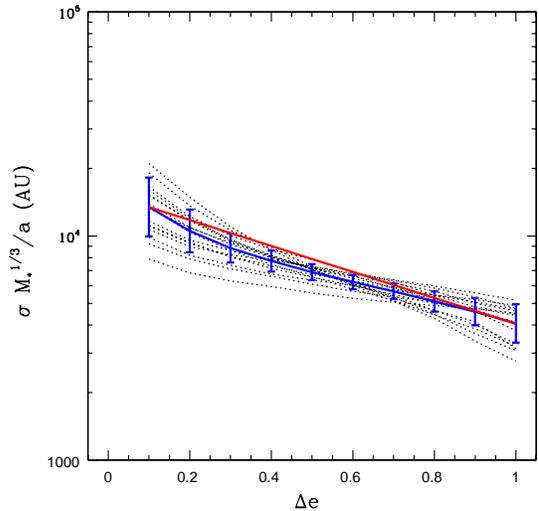,width=8.5cm}} 
\caption{Scaled cross sections versus eccentricity increase $\Delta{e}$
(equivalently, post-encounter eccentricity $e$) for solar systems with 
different stellar masses.  The starting state is taken to have four
giant planets with the current masses and semimajor axes.  Cross
sections are scaled by $M_\ast^{1/3}/a$ (see text).  The individual
cases are shown as light dotted curves, with include curves for each
of the four planets for four choices of stellar mass $M_\ast$ = 0.25
-- 2.0 $M_\odot$ (spaced by factors of 2).  The heavy solid blue curve
depicts the average, where the error bars depict the standard
deviation. The red striaght line shows the result for cross sections
with a purely exponential dependence on eccentricity. }
\label{fig:mscale} 
\end{figure} 

The dependence of the interaction cross sections on the mass of the
host star is somewhat more complicated than for the other variables,
as illustrated in Figure \ref{fig:mcross}. As the mass $M_\ast$ of the
star increases, the cross sections, considered as functions of
eccentricity, become steeper. The spacing of the curves in Figure
\ref{fig:mcross} (for different stellar masses) grows with $e$, so
that the curves are not self-similar. In spite of this complication,
we can still fit the cross sections with a power-law function of
stellar mass, although the accuracy of the approximation is not
expected to be as high as in the previous cases. We thus consider a
scaling of the form $\cross \propto M_\ast^{-\mu}$, and vary the index
$\mu$ to find the best fit. The choice $\mu=1/3$ provides the lowest
RMS of the relative error. Figure \ref{fig:mscale} shows the result by
plotting the scaled cross sections $\cross M_\ast^{1/3}/a$ (again
including the linear dependence on semimajor axis $a$) as a function
of post-encounter eccentricity.  The light dotted lines show the
individual (scaled) cross sections and the heavy blue curve shows the
average. The error bars depict the corresponding standard deviation,
which is larger than for the cases considered previously (compare
Figure \ref{fig:mscale} with Figures \ref{fig:scale} and
\ref{fig:vscale}). The heavy red staight line shows the same result as
before (from equation [\ref{expeccent}]).  The RMS error between the
exponential line and the numerically determined cross sections is about
20\%.  This larger error measure results from fitting the cross
sections with a power-law form, even though the results depart somewhat 
more from self-similarity. 

\begin{figure}  
\centerline{\psfig{figure=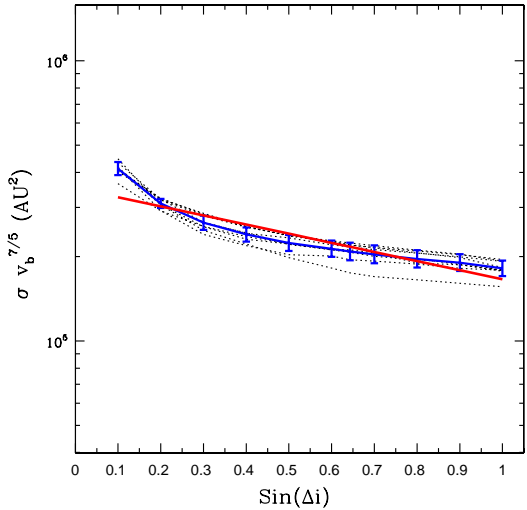,width=8.5cm}} 
\caption{Scaled cross sections for increasing the post-encounter  
spread $\Delta i$ of the inclination angles of the planetary orbits.
The starting states have the four giant planets orbiting in the same
plane ($\Delta i=0$). The cross sections are scaled by the velocity
dispersion of the cluster with the relation $\cross \vdisp^{7/5}$.  
The individual cases are shown as light dotted curves. The heavy 
solid blue curve depicts the average, whereas the error bars depict 
the standard deviation. The heavy red curve shows the fitting 
function described in the text. }
\label{fig:iscale} 
\end{figure} 

The cross sections for increasing the spread of inclination angles,
considered over a range of velocity dispersions, also show a nearly
self-similar form (see Figure \ref{fig:icross}).  This finding
indicates that the cross section should scale with a nearly power-law
dependence so that $\cross \propto \vdisp^{-\eta}$.  Over the range
$\vdisp=1-16$ km/s, we find that the best fit occurs for
$\eta\approx7/5$. To illustrate how well this scaling law works, we
plot the scaled cross sections $\cross \vdisp^{7/5}$ as a function of
$\sin(\Delta i)$ in Figure \ref{fig:iscale}. Each light dotted curve
in the figure corresponds to the result of one choice of velocity
dispersion.  The heavy blue curve shows the average of the scaled
cross sections, where the error bars depict the standard deviations. 
The mean size of the error bars corresponds to relative differences 
of $\sim6\%$, so that the curves are self-similar to this degree of 
accuracy.  Notice that the scaling exponent $\eta\approx7/5$ for
inclination angle increases as a function of velocity dispersion
$\vdisp$ is the same as the corresponding index for eccentricity
increases.

After the velocity dependence has been scaled out, the cross section
for increasing the spread of inclination angles is a slowly varying
monotonic function of $\Delta i$ (see Figure \ref{fig:iscale}). If we
consider $x=\sin\Delta{i}$ as the independent variable (instead of
$\Delta i$ itself), the cross section can be fit with an exponential
function which is analogous to that used to describe the eccentricity
dependence.  More specifically, if we use the functional form 
\be
\cross_i = \cross_0 \exp \left[ 
b_0 \left( 1 - \sin\Delta{i} \right) \right] \,,
\label{expangle} 
\ee
then the cross section for increasing $\Delta i$ can be fit using the
parameters $b\approx3/4$ and $\sigma_0\approx166,000$ AU$^2$. Note
that the value of the index $b$ used here somewhat smaller than that
needed to fit the dependence of the cross sections on (post-encounter)
eccentricity (compare with equation [\ref{expeccent}]). The fitting
function from equation (\ref{expangle}) is shown in Figure
\ref{fig:iscale} as the solid red curve. The quality of the fit is
reasonably good: The fitting curve falls within one standard deviation
(marked by errorbars in the figure) of the mean for all of the range
except the first point ($x=\sin\Delta{i}=0.1$); alternately, the RMS of
the relative error between the two curves is $\sim8\%$. However, the 
mean of the numerical results (blue curve) shows more curvature than 
the exponential fit (red curve), especially at small values of $x$. 

Although we could find a more complicated fitting function that has
smaller RMS relative error, we use equation (\ref{expangle}) in order
to compare changes in the spread of inclination angle with changes
orbital eccentricity.  If we equate the variable $x=\sin\Delta{i}$
with $e$, then equations (\ref{expeccent}) and (\ref{expangle}) have
the same general form.  We can then compare the leading coefficients,
which have values $\cross_0\approx$ 166,000 AU$^2$ for
$\Delta{i}$-dependence and $\alpha\pi\ell{a}\approx$ 120,000 AU$^2$
for $e$-dependence, where we have used $a$ = 30 AU to evaluate the
latter expression. The cross sections for eccentricity increase and
spread of the inclination angles thus display similar behavior. The
leading coefficients agree to within $\sim28\%$ and we can make the
following inexact analogy: An increase in Neptune's eccentricity of
$\Delta e$ = 0.1 corresponds to changing the spread of the inclination
angles (of all four planets) so that $\sin\Delta{i}$ increases by
0.1. We can also make a rough association between increasing the
spread of inclination angles to $\Delta i \ge 90$ degrees and the
ejection of a planet ($e\ge1$). Both of these events have
(approximately) the same cross section and both involve order unity
changes to the angular momenta of the planetary orbits. In addition,
the cross sections for inclination angle increase and eccentricity
increase scale with the velocity dispersion in the same manner
($\propto \vdisp^{7/5}$).

The association between changes in the variables $\sin\Delta{i}$ and
${e}$ provides an intriguing topic for additional work. To leading
order, the canonical actions written in terms of the orbital elements
have the forms 
\be
\Gamma \propto {1 \over 2} e^2 \qquad {\rm and} \qquad 
Z \propto \sin^2(i/2) \,.
\ee
The apparent relation between the two variables (as observed in the
simulation results) could thus be evidence of an equipartition-like
mixing of the actions (see \citealt{lichtlieb}). Although beyond 
the scope of the present paper, this issue should be explored further.  

We can extract a potentially important cross section from these
results.  The scattering interactions considered here can readily
increase the spread of inclination angles of outer bodies in a solar
system. On the other hand, the scattering events themselves have
little effect on planets in tight orbits, such as the multi-planet
systems observed by the {\it Kepler} mission \citep{batkepler}.
However, the bodies in the outer solar system can have important
long-term effects on the inner bodies provided that they are scattered
into orbits with sufficiently high inclination angles. More
specifically, if the inclination angles of the outer orbits are larger
than 39.2$^\circ$, then the Kozai effect can operate \citep{kozai,lidov}, 
and the inner portions of the solar system can be excited over the age
of the systems. Combining this requirement with the results of our
numerical simulations, we find that the cross section for scattering a
solar system into a state where the Kozai effect can operate is given by 
\be
\cross_{\rm kozai} \approx 210,000 \,{\rm AU}^2\,
\left( {a_{\rm out} \over 30\,{\rm AU}} \right) 
\left( {\vdisp \over 1\,{\rm km/s}} \right)^{-7/5} \,, 
\label{kozaicross} 
\ee 
where $a_{\rm out}$ is the semimajor axis of the outermost planet of
the system. Note that the requirement of large mutual inclination is
necessary but not sufficient for the Kozai effect to play a role. 
The Kozai effect is a highly fragile type of interaction because it
involves libration of the argument of periastron, and this quantity
can be subject to many other sources of precession (for further
discussion, see \citealt{bmt}).  We also note that this form for the
cross section (equation [\ref{kozaicross}]) involves some
extrapolation: The numerical simulations were carried out primarily
for the architecture of the current solar system.  Nonetheless, the
outermost planet is always the most affected by fly-by interactions, 
and the cross sections scale linearly with semimajor axis to a good
approximation.

\begin{figure} 
\centerline{\psfig{figure=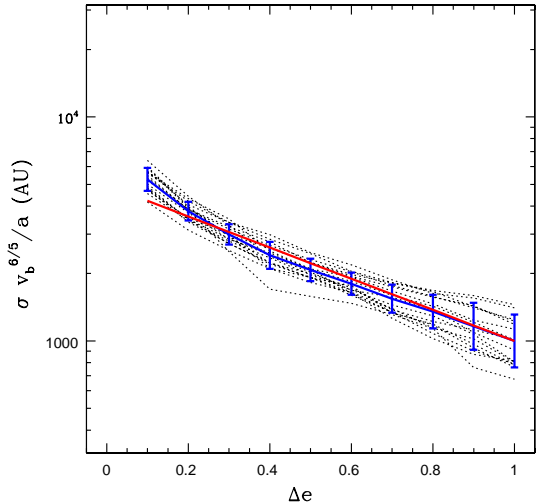,width=8.5cm}} 
\caption{Scaled cross sections versus eccentricity increase $\Delta{e}$
(equivalently, post-encounter eccentricity $e$) for solar systems 
interacting with single stars. The starting states have four giant
planets with the current masses and semimajor axes of our Solar
System.  Cross sections are scaled by the factor $\vdisp^{6/5}/a$.
The individual cases are shown as light dotted curves, which include
the four giant planets and four values of velocity dispersion of the
background cluster: $\vdisp$ = 1, 2, 4, and 8 km/s. The heavy solid
blue curve depicts the average, where the error bars depict the
standard deviation. The red striaght line shows the result for cross
sections with an exponential dependence on eccentricity. }
\label{fig:singlescale} 
\end{figure}  

Next we consider the scaling behavior of the cross sections for
interactions with passing single stars. As for the case of binary
systems, we expect the cross sections to scale nearly linearly with
the semimajor axis $a$ of a given planet. In addition, the nearly
equal spacing on the logarithmic plot of Figure \ref{fig:singlecross}
indicates that the cross sections should display power-law dependence
on the velocity dispersion, such that $\cross\propto\vdisp^{-\gamma_s}$.  
The velocity dependence for these single star cross sections is
moderately less steep than those found earlier for binaries; the
optimal value of the index $\gamma_s\approx6/5$, which is somewhat
smaller than the value for binary cross sections $\gamma\approx7/5$. 
After scaling out the semimajor axis and velocity dispersion, the
reduced cross sections are shown in Figure \ref{fig:singlescale}. 
The light dotted curves show the scaled values for given planets and
values of $\vdisp$ (which lie in the range 1 -- 8 km/s). The heavy
blue curve shows the mean over the entire collection and the error
bars denote the standard deviations. These error bars correspond to an
average relative error of $\sim15\%$, which is comparable to, but
somewhat larger than that found for the binary cross sections.

The scaled cross sections shown in Figure \ref{fig:singlescale} for
single star interactions show a nearly exponential dependence on the
post-encounter eccentricity. Although this behavior is analogous to
that found for the binary cross sections, the slope of the exponential
is somewhat steeper. Here we consider a fitting function of the form 
\be
\cross_{\rm single} = \cross_0 \, \left( {a \over {\rm AU}} \right) 
\left( {\vdisp \over 1\,{\rm km/s}} \right)^{6/5} 
\exp \left[ b_s (1-e) \right], 
\label{singlefit} 
\ee
where we obtain a good fit for $\cross_0=1000$ AU$^2$ and $b_s=8/5$.
The resulting fit is shown as the red straight line in Figure
\ref{fig:singlescale}. The RMS difference between the expression of
equation (\ref{singlefit}) and the numerically determined, scaled
cross sections for single stars is only $\sim8\%$.

Now we can compare the cross sections for single stars with those for
binaries. The comparison is complicated by the different scalings of
the two cases with velocity dispersion and the different exponential
laws for the eccentricity dependence. To fix ideas, consider the
benchmark case where the velocity dispersion $\vdisp$ = 1 km/s for the
background cluster. Here, the cross sections for binary star
interactions are $\sim4.2$ times larger than those for single stars at
the high end of the eccentricity range $e=1$. Similarly, the binary
cross sections are $\sim3.2$ times larger at the low end of the
eccentricity range where $e=0.10$.  Averaged over the span of
eccentricity considered here, the binary cross sections are larger by
a factor of $\sim3.6$.  This factor decreases with increasing velocity
dispersion, however, because the binary cross sections fall according
to the relation $\cross\propto\vdisp^{-7/5}$, whereas the single star
cross sections fall as $\cross\propto\vdisp^{-6/5}$. With these
scaling laws, the cross sections for binaries are only a factor of 2
larger (than those for single stars) when the velocity dispersion is
increased to $\vdisp\approx20$ km/s. 

These results can be interpreted as follows: At high asymptotic
speeds, which occur for $\vdisp\gta20$ km/s, the two members of a
binary pass by the solar system quickly enough so that binary motion
and planetary motion play only a minor role in the interaction (this
speed is much larger than the mean orbital speed of either the binary
or the outer planet).  As a result, the two stars interact with the
solar system in an almost independent manner, and the cross sections
for binary interactions should be a factor of $\sim2$ larger than
those for single stars (for large $\vdisp$). On the other hand, lower
impact speeds can be comparable to the binary orbital speed and/or the
planetary orbital speeds. In this regime, the motion of the binary
stars relative to one another during the encounter can increase their
chances of interacting with the planets, thereby leading to larger
cross sections. In extreme cases, resonant interactions can occur when
the velocity scales of the problem are all comparable (see also
\citealt{frozen}), and these long-lived events can greatly increase
the chances of disruption of planetary orbits during the encounters.
To be consistent with this picture, the ratio of the single-star cross
section to the binary cross section must decrease less steeply with
$\vdisp$, as found here. 

Before leaving this section, we briefly address the issue of how
self-similarity can arise in the context of solar system scattering.
In its full form, this problem has six velocities (four planetary
orbits, one binary orbit, and the encounter velocity) and seven masses
(four planets and three stars). One expects self-similarity only when
most of these scales do not contribute \citep{barenblatt}. We can
construct an argument to reduce the number of relevant scales as
follows: To leading order --- and only during the encounter itself ---
planetary interactions with the binary are independent of interactions
with other planets.  As a result, we can (often) treat the encounters
as single-planet systems scattering with binaries. The planet itself
is usually small enough to be considered as a test mass, so that we
are left with ``only'' three masses and three velocities. The binary
masses are always drawn from the same IMF, and the cross sections are
determined through many samples of that IMF (${\cal N}_E\gta80,000$),
thereby leaving the ratio $M_\ast/(\mone+\mtwo)$ as the most important
mass variable. In the regime of interest, the cross sections have
values in the range $\cross\sim10^4-few\times10^5$ AU$^2$, which
implies that the length scales that characterize the interactions
$\ell_{\rm c}\equiv\cross^{1/2}\approx100-500$ AU. This size scale is 
larger than that of both the planetary orbits ($a=5-30$ AU) and most
binary orbits (where the peak of the period distributions corresponds
to $a_{\rm b}\approx42$ AU). If the orbital speeds of the planets and
the binary are fast enough, then their orbits can be replaced by rings
of mass with the same semimajor axis and eccentricity \citep{md1999}.
This averaging effectively eliminates the orbital velocities from the
problem and leaves the velocity dispersion $\vdisp$ as the most
important velocity variable. Indeed, we find that the cross sections
depend most sensitively on the stellar host mass $M_\ast$
(equivalently, the mass ratio $M_\ast/(\mone+\mtwo)$) and the velocity
dispersion $\vdisp$. This argument is not exact, however, and the
additional scales (e.g., orbits speeds) do play some role. These
complications are responsible for the spread in the scaled cross
sections shown in Figures \ref{fig:clinear} -- \ref{fig:iscale}.

We can also compare these scaling results to analytic results found in
previous studies (see, e.g., \citealt{heggie,spurzem2009}), although
the system parameters are not exactly the same. The latter study finds
a scaling relation $\cross\propto a^{3/2}\vdisp^{-1}$ in the impulsive
regime (where $v_\infty\sim\vdisp$ is much greater than the orbital
speed of the planet) and $\cross\propto a\vdisp^{-2}$ for
non-impulsive encounters. Our results (see Figures \ref{fig:vscale}
and \ref{fig:singlescale}) are intermediate between these two scaling
laws, since the encounters are rarely fully in the impulsive or the
non-impulsive regime. In addition, this current study includes
binaries, and the binary orbital speed is generally comparable to the
planetary orbital speed. The binary motion can either add to or
subtract from the relative velocity of the encounter (depending on the
timing and geometry of the encounters), so that the scattering
interactions have a wide range of relative velocities, even for a
given $\vdisp$. As a result, our parameter space does not fall fully
in any of the limiting regimes considered by previous analytic
estimates.

\section{The Solar Birth Aggregate} 
\label{sec:application}

Given that most stars are born within clusters, it is likely that the
birth environment of our own Solar System was a cluster of some type.
The argument for a substantial birth cluster is bolstered by evidence
for short-lived radionuclides in meteorites, which suggests that the
early solar nebula was enriched by a nearby supernova (\citealt{bigal}; 
see the review of \citealt{dauphas}). A number of previous papers have
considered how dynamical scattering encounters in this putative birth
cluster can provide constraints on the cluster properties (see the
discussion of Section \ref{sec:intro}). Unfortunately, however, no
consensus has been reached. This section briefly revisits the issue in
light of the updated cross sections determined above.

The basic problem posed by the solar birth aggregate involves a number
of ingredients: {[I]} Direct supernova enrichment of the early solar
nebula requires a nearby massive star, which is more likely to form in
a larger stellar system. Further, significant nuclear enrichment
requires close proximity (distances $d = 0.1 - 0.3$ pc), which implies
that the supernova progenitor lives within the same cluster. Acting in
the opposite direction, larger clusters can potentially disrupt
planetary systems through the action of both {[II]} dynamical
scattering (with the cross sections determined here) and through
{[III]} intense radiation fields which can evaporate gaseous disks. In
order for the solar system to reach its present-day state, however,
the orbits of the giant planets cannot be greatly perturbed and the
early solar nebula could not be too severely evaporated. On the other
hand, {[IV]} the classical Kuiper belt has an apparent edge at
$\sim50$ AU, and {[V]} the dwarf planet Sedna has an unusual orbit;
both of these solar system properties could be explained by 
{\it requiring} a close encounter with another member of the cluster.
The challenge is to find a birth scenario for the solar system that
successfully negotiates the compromises required to simultaneously
explain all five of these constraints. Supernova enrichment, the edge
of the Kuiper belt, and the orbit of Sedna all argue in favor a large
and long-lived cluster; disruption via both scattering and radiation
argue in the opposite direction.

Existing work has considered a variety of approaches to this issue.
Several authors advocate solar birth clusters with stellar membership
size in the range $N=10^3-10^4$ (e.g., \citealt{adams2001,zwart};
\citealt{adams2010,pfalzner}). These studies find that cluster systems
in this decade of $N$ lead to moderate dynamical disruption of their
constituent planetary systems. Additional work focuses on even larger,
longer-lived clusters and find that they can instigate substantial
changes to planteary orbits, including frequent ejections
\citep{malmberg2007,malmberg2011,spurzem2009,parker2012,hao2013}. On
the other hand, competing work suggests that the solar birth cluster
does not produce significant disruption of planetary orbits
(\citealt{willgaidos,dukes,craig}; see also \citealt{williams2010}).

The aforementioned papers thus reach different conclusions about the
importance of dynamical scattering of planetary systems in clusters.
These differences arise because of varying assumptions about cluster
properties and varying assumptions about how to enforce the five
constraints on solar system properties outlined above. Although a full
review of this topic is beyond the scope of this work, we provide a
brief overview below (for additional detail, see the reviews of
\citealt{adams2010,dauphas,pfalzner}).

For a given type of disruption, with cross section $\cross$, the
interaction rate is given by $\Gamma=n_\ast\cross{v}$ (from equation
[\ref{rate}]). The total expected number $N_{\rm dis}$ of disruption
events, per solar system, integrated over the lifetime $\tau$ of the
cluster is then given by 
\be
N_{\rm dis} = \int_0^\tau \Gamma dt = 
\int_0^\tau n_\ast \cross v dt \,. 
\label{ndisrupt} 
\ee 
The number of disruptive interactions thus depends on the speed $v$ at
which a given solar system encounters passing binaries, their number
density $n_\ast$, and the total time $\tau$ spent within the cluster.

We first consider the speed $v$. Recall that the interaction cross
section $\cross$ varies with the velocity dispersion of the cluster
according to the relation $\cross\propto\cross_0\vdisp^{-7/5}$. If we
identify the speed $v$ with the velocity dispersion $\vdisp$ of the
cluster, then the product $\cross v \propto \vdisp^{-2/5}$. As a
result, most of the velocity dependence of the cross section is
compensated by that of the interaction rate, so that the number of
disruption events depends only weakly on the velocity dispersion.
As an example, consider the Orion Nebula Cluster (ONC), an
intermediate-sized young stellar system with velocity dispersion
$\vdisp\sim2$ km/s \citep{hillhart}. Provided that it stays intact,
the ONC is likely to evolve into an open cluster resembling the
Pleiades \citep{kroupa2001}; over the coming $\sim100$ Myr, the
velocity dispersion of the cluster will slowly decrease to
$\vdisp\sim1$ km/s. Over this span of time, the quantity
$\vdisp^{-2/5}$ that defines the velocity dependence of the
interaction rate varies by only about 32\%.

For setting the number of disruption events, one important quantity 
is the time $\tau$ over which clusters remain intact as dynamical
systems. In the simplest terms, although most stars are formed in
clusters, these astronomical entities come in (at least) two
distinctly different flavors. Only about 10 percent of the stellar
population is born within clusters that are sufficently robust to
become open clusters \citep{roberts,bc1991}, which are relatively
long-lived ($\tau$ = 100 Myr -- 1 Gyr). The remaining 90 percent of
the stellar population is born within embedded clusters (e.g.,
\citealt{allenppv}), which have much shorter lifetimes ($\tau\sim10$
Myr). As shown below, solar systems that are born within long-lived
clusters can have an appreciable chance of dynamical disruption;
short-lived clusters lead to significant disruption with greatly
reduced probability.

Another important quantity is the density of the cluster. For clusters
found in the solar neighborhood, the cluster radius $R\propto N^{1/2}$
\citep{ladalada}, so that the clusters display nearly constant surface
density \citep{adams2006}. With this relation, clusters with larger
stellar membership sizes $N$ have lower mean densities. However, the
clusters in the sample are relatively small (with $N < 2500$), and
this trend does not continue up to the largest clusters with
$N=10^4-10^6$ \citep{whitmore}, or to the subpopulation of systems
that become globular clusters. The largest clusters can thus have
larger densities.

To assess the effects of scattering encounters, we need to specify the
rate $\Gamma$ at which solar systems encounter passing binaries (and
single stars). As shown previously \citep{adams2006,proszkow2009}, the
rate of close encounters in a cluster can be written in the convenient
form 
\be
\Gamma = \Gamma_0 \left( {b \over b_0} \right)^\gamma \,,
\ee
where $b_0$ is a fiducial distance (taken here to be $b_0$ = 1000 AU), 
and where the fiducial rate $\Gamma_0$ and index $\gamma$ depend on the
cluster properties. The index $\gamma$ falls in the range
$1\le\gamma\le2$, where the extreme of the range correspond to perfect
gravitational focusing ($\gamma\to1$) and the full geometrical cross
section ($\gamma\to2$). In these systems, encounters beyond $\sim1000$ AU 
are little affected by gravitational focusing. Since the cross sections 
calculated in this paper include gravitational focusing, we can write 
the interaction rate in the form 
\be
\Gamma = \Gamma_0 {\cross \over \pi b_0^2} \,.
\ee
The benchmark interaction rate $\Gamma_0$ has a typical value of about
0.1 interactions per target star per Myr. However, given the wide
range of possible cluster properties, it can vary over a wide range,
from an order of magnitude lower to an order of magnitude larger than
this fiducial value (see Figures 6 and 7, and Tables 8 -- 13 in
\citealt{proszkow2009}). Note that the benchmark rate is, in general,
larger than the simple estimate $\Gamma_0\sim\nbar\vdisp\pi{b_0^2}$,
where $\nbar$ is the mean density of the cluster. The stellar density
that defines the interaction rate is not the mean over the cluster,
but rather the weighted mean over the integrated orbits of the
ensemble of cluster members. The cluster members generally do not stay
at a given cluster radius, and the cluster density is centrally
concentrated, so that solar systems sample the higher stellar
densities of the cluster core. This effect is amplified by the
starting conditions for clusters, which start with subvirial initial 
conditions; as a result, the orbits are more radial than isotropic, 
resulting in more excursions through the dense central core (see 
\citealt{adams2006,proszkow2009} for further discussion). 

Collecting the results outlined above, we can write the number of
disruption events (from equation [\ref{ndisrupt}]) in the form
\be
N_{\rm dis} \approx {\cross_0 \over \pi b_0^2} \int_0^\tau 
\Gamma_0 \left( {\vdisp \over 1\,{\rm km/s}} \right)^{-2/5} dt\,. 
\label{ndtwo} 
\ee
The cross section for moderate solar system disruption can be taken as
$\cross_0\approx2.5\times10^5$ AU$^2$, which corresponds to events
producing eccentricity increases $\Delta e$ = 0.1 and/or increases in
the spread of inclination angles $\Delta i = 10^\circ$ (e.g., see 
Figures \ref{fig:solarcom} and \ref{fig:icross}). To obtain this
value, we use a linear combination of the binary and single-star cross
section (see Figure \ref{fig:singlecross}), and an assumed binary
faction of 2/3. Although these changes to the orbital elements are not
devastating, they are large enough to distinguish a disrupted solar
system from our own. Note that this value can be written
$\cross_0\approx(500\,{\rm AU})^2$, which is somewhat larger than the
previous estimate of $\sim(400\,{\rm AU})^2$ (from
\citealt{adams2001}).\footnote{The difference arises because the
  present study increases the target area in equation (\ref{crossdef})
  from $B=4$ to $B=100$, thereby including more distant events. Note
  that the original work \citep{adams2001} correctly introduced the
  cross sections as lower limits. The present cross sections are also
  lower limits, although they are much closer to their greatest lower
  bounds.}  The leading factor in equation (\ref{ndtwo}) is thus of
order 1/10. Since the benchmark interaction rate $\Gamma_0\sim0.1$
Myr$^{-1}$, the cluster lifetime must be relatively long,
$\tau\sim100$ Myr, in order for disruption to take place with high
probability. In other words, most solar systems residing in long-lived
clusters can experience moderate disruption.

The cross sections for planet ejection are smaller than the values
used above by a factor of $\sim3$. As a result, only a fraction
($\sim1/3$) of the solar systems in long-lived clusters are expected
to lose planets with Neptune-like orbits (with even smaller fractions
for closer planets). Keep in mind, however, that the benchmark
interaction rates $\Gamma_0$ can vary by a factor of $\sim10$ in both
directions.

The above considerations resolve some of the differences found in the
literature concerning the disruption rates for planetary systems in
clusters. In order for disruption to occur with high probability,
clusters must live for relatively long times $\tau \gta 100$ Myr.
Indeed, the studies that find low disruption rates consider the
clusters to have relatively short lifetimes $\tau \sim 10$ Myr (e.g.,
\citealt{willgaidos,dukes}).

How long are clusters expected to stay together? As outlined above,
the cluster population has at least two branches. Some clusters
disperse over relatively short time scales of only $\sim10$ Myr. The
robust clusters that survive to become open clusters have empirically
determined lifetimes $\tau_{em}$ that can fit with a function of the
form
\be
\tau_{em} = 2.3 {\rm Myr} 
\left( {M_c \over 1 M_\odot} \right)^{0.6} \,,
\label{lifetime} 
\ee
where $M_c$ is the cluster mass \citep{lamers}. With this relation,
clusters with initial masses larger than $\sim550M_\odot$ live longer
than 100 Myr and can potentially disrupt their constituent solar
systems. More specicifally, we can write the dynamical constraint 
in the form 
\be
N_{\rm dis} \approx {\cross_0 \over \pi b_0^2} 
\langle \Gamma_0 \rangle 2.3 (\Upsilon N)^{3/5} 
\langle (\vdisp/1\,{\rm km/s})^{-2/5} \rangle \lta 1 \,, 
\ee
where $\Upsilon$ ($\approx1/2$ $M_\odot$/star) is the mass-to-number
ratio (the conversion factor between cluster mass $M_c$ and cluster
membership size $N$), and where we include the time average of the
velocity dispersion of the cluster (raised to the proper power).  
After some rearrangement and the specification of typical numbers,
this constraint can be written in the form 
\be
N \lta 5000 \Bigl[ 
\left( {\cross_0 \over 2.5 \times 10^5 {\rm AU}^2} \right) 
\left( {\langle \Gamma_0 \rangle \over 0.05 {\rm Myr}^{-1}} \right)
\label{constraint} 
\ee
$$
\qquad \times \, 
\left\langle \left( {\vdisp \over 1\,{\rm km/s}} \right)^{-2/5} \right\rangle
\Bigr]^{-5/3} \lta 10^4 \,.  
$$ 
Note that the disruption cross section is determined more precisely
than either the expected age of the cluster (from equation
[\ref{lifetime}]) or the benchmark interaction rate $\Gamma_0$. This
latter quantity can be determined to high accuracy for a given set of
cluster properties and initial conditions, but its value varies
appreciably from cluster to cluster \citep{proszkow2009}. Equation
(\ref{constraint}) uses a value near the low end of the range in order
to provide an upper limit on $N$.  In light of these uncertainties, a
reasonable order-of-magntidue estimate for the dynamical constraint is
$N\lta10^4$, as given by the final inequality. This result is roughly
consistent with previous estimates \citep{adams2001,zwart,adams2010,pfalzner}. 
Nonetheless, the full probability distribution for the survival (or
disruption) of planetary systems as a function of cluster size $N$
should be constructed.

The constraint given by equation (\ref{constraint}) assumes that the
solar birth cluster is relatively long-lived. If the solar system
formed within a cluster that disperses in only $\sim10$ Myr, the
corresponding dynamical constraint would be considerably weaker. The
motivation for considering a long-lived cluster comes from constraints
jointly implied by the five solar system properties outlined at the
beginning of this section. Direct supernova enrichment [I] favors a
long-lived cluster, so that the progenitor star has enough time to
live, evolve, and explode. An even stonger argument comes from the
need for a scattering event to produce the edge of the classical
Kuiper belt [IV] and to produce the orbit of Sedna [V]. If these solar
system properties arise from dynamical interactions in the birth
cluster, then a long-lived stellar system is strongly indicated. It
remains possible for these features of the solar system to be
explained in other ways. Nonetheless, any self-consistent set of
constraints on the solar birth environment must explain all three of
these properties, and must simultaneously account for the
corresponding constraints due to dynamical scattering encounters [II]
and radiation fields [III] (e.g., see \citealt{fatuzzo,thompson}).

For completeness, we also consider possible constraints on the solar
birth cluster for the scenario where the solar system spends much of
its early life in the ultra-compact multi-resonant configuration (see
Section \ref{sec:results}, Figure \ref{fig:solarres}).  The cross
section for removing the solar system from its resonant state is then
given by equation (\ref{rescross}), which is more than nine times
larger than that used above. If, in addition, the removal of the solar
system from resonance {\it always} led to significant disruption over
longer times, then the maximum size of the solar birth cluster would
be $\sim40$ times smaller than that of equation (\ref{constraint}).
In practice, however, the solar system, after being removed from
resonance, will not always be significantly disrupted (for example by
ejecting a planet) before it evolves and spreads out (as advocated by
the Nice model; \citealt{gomes2005,tsiganis}). To assess the risk of
disruption in this case, one must also know the probability of the
non-resonant (but still compact) solar system experiencing disruption
on sufficiently short time scales. This calculation involves a large
ensemble of long-term ($\sim100$ Myr) solar system integrations and is
beyond the scope of this present work. Nonetheless, direct application
of equations (\ref{rescross}) and (\ref{constraint}) suggests that the
constraint could be tighter than that derived for the solar system in
its usual configuration.

Finally, we note that another class of observational constraints on
the solar birth environment might become available. Given that the
birth cluster is expected to have $N\approx10^3-10^4$ stars with
similar chemical composition, it is possible in principle to find
other members of our solar birth aggregate. Although billions of years
have passed and the cluster has long since dispersed, perhaps $\sim20$
of these solar siblings could reside within 100 pc of the Sun
\citep{zwart}. By focusing on the chemical species that show the most
variation from cluster to cluster, it is possible to observationally
distinguish these siblings from other stars \citep{ramirez}. The
discovery of even a few such stars would provide strong constraints on
the properties of the solar birth cluster and its location within the
Galaxy. On the other hand, the Solar System could have had a more
complicated dynamical history including large radial migration in the
Galaxy \citep{kaib}, which could reduce the chances of finding solar
siblings. 

\section{Conclusion} 
\label{sec:conclude} 

\subsection{Summary of Results} 
\label{sec:summary} 

Using results from more than 2 million individual numerical scattering
experiments, this paper has found cross sections for the disruption of
planetary orbits in solar systems interacting with passing stars and
binaries. Our specific results can be summarized as follows:
 
[1] More compact solar systems have smaller interaction cross sections
(Figures \ref{fig:solarcom} and \ref{fig:solarres}).  To leading
order, the cross section for a given disruption event (e.g., planet
ejection or eccentricity increase) scales linearly with the semimajor
axis of the initial orbit, i.e., $\cross \propto a$ (see Figure
\ref{fig:clinear}).

[2] For most solar systems, the cross section for a given planetary
orbit to be disrupted during a scattering encounter is almost
independent of the other planets. This feature of the interactions
allows for the scaling analysis presented in Section
\ref{sec:analysis}.  Of course, after the encounter, solar systems
that suffer moderate disruption can subsequently experience orbital
instability, and this latter effect does depend (quite sensitively) on
the other planets in the system. In addition, for highly
self-interacting solar systems, those with sufficiently massive
planets and/or close orbits, interactions among the planets themselves
can lead to effectively higher cross sections (e.g., see Figures
\ref{fig:solarres} and \ref{fig:fatcross}).

[3] The dependence of the cross sections $\cross$ on the
post-encounter eccentricity $e$ has a nearly exponential form (see
Figures \ref{fig:solarcom} -- \ref{fig:efinite}).  As a result, the
cross sections can be written $\cross \propto \exp[-be]$, where $b
\approx 4/3$ provides a good fit across the range of parameter space
considered in this work (Figures \ref{fig:scale}, \ref{fig:vscale},
and \ref{fig:mscale}).

[4] The cross sections depend sensitively on the velocity dispersion
$\vdisp$ of the background environment, where the dependence displays
a nearly power-law form. Moreover, the shape of the cross section
curves, as a function of eccentricity, are nearly the same across the
parameter space considered here (Figures \ref{fig:vcross} and
\ref{fig:vscale}).  The cross sections can thus be written as
$\cross\propto\vdisp^{-\gamma} \exp[-be]$, where $\gamma$ = 7/5 and
$b$ = 4/3 provide a good fit over range of interest.

[5] The cross sections depend on the mass $M_\ast$ of the host star,
where the dependence has the approximate form $\cross \propto
M_\ast^{-1/3}$. The mass dependence is somewhat more complicated,
however, as the cross sections are not fully self-similar (see Figures
\ref{fig:mcross} and \ref{fig:mscale}). For more disruptive encounters
(where $e\to1$ and planets are ejected), the scaling with mass is
somewhat steeper and the form $\cross \propto M_\ast^{-1/2}$ provides
a better fit (consistent with previous results from
\citealt{adams2006}).

[6] Most of this work considers planetary orbits with vanishing
initial eccentricity $e$. Nonetheless, for solar systems starting with
$e\ne0$, the interaction cross sections for eccentricity increase are
nearly the same (Figure \ref{fig:efinite}), provided that one
considers post-encounter eccentricities sufficiently larger than the
starting values (roughly, by the increment $\delta e \sim 0.1$). This
finding stands in contrast to the related problem of single stars
interacting with binaries, where the cross sections for binaries with
$e=0$ and $e\ne0$ are significantly different \citep{heggie}.

[7] The above results can be combined to write the cross section for
eccentricity increase for solar systems interacting with binaries in
the general form 
\be
\cross = 4050 \,\, ({\rm AU})^2 \,\,
\left( {a \over {\rm AU}} \right) \,
\left( {M_\ast \over M_\odot} \right)^{-1/3} \,
\label{universal} 
\ee 
$$
\qquad \times 
\left( {\vdisp \over 1 {\rm km/s}} \right)^{-7/5}\,
\exp\left[ {4 \over 3} \left(1-e\right)\right]\,.  
$$ 
This result holds over the ranges of parameters given by 5 AU
$\le{a}\le$ 50 AU, $0.25M_\odot\le{M_\ast}\le2M_\odot$, 1 km/s
$\le\vdisp\le16$ km/s, and $0.1\le{e}\le1$. Equation (\ref{universal})
is in good agreement with the numerically obtained results: For fixed
stellar mass, the RMS relative error for the range of starting
semimajor axis, velocity dispersion, and post-encounter eccentricity
is less than about 12 percent (see Figures \ref{fig:scale} and
\ref{fig:vscale}). Including variations in the stellar mass, the RMS
error is less than about 20 percent (Figure \ref{fig:mscale}). Over
the same regime of parameter space, the cross section itself varies by
more than a factor of $\sim1000$. Equation (\ref{universal}) provides 
the total ejection cross sections (including capture events) in the 
limit $e\to1$; the cross sections for ejection and capture are listed 
separately in Table 1.  

[8] The cross sections for increasing the spread of inclination angles
$\Delta i$ are comparable to those for increasing eccentricity (Figure
\ref{fig:icross}). The cross sections for $\Delta i$ also show a
nearly self-similar form, and scale with velocity dispersion of the
background cluster according to $\cross \propto \vdisp^{-7/5}$ (Figure
\ref{fig:iscale}). This scaling exponent is the same as that found for
eccentricity increases. The cross sections can be fit with an
exponential dependence on the variable $x=\sin\Delta{i}$. Although
inexact, one can identify increases in inclination with increases in
eccentricity such that $\Delta{x}\sim\Delta{e}$. In general, increases
in the spread of inclination angles and orbital eccentricity are
well-correlated (Figure \ref{fig:ive}), although the $\Delta{i}$
values for a given $\Delta{e}$ display a wide range. We have also
determined the cross sections for increasing the inclination angles
beyond 39.2$^\circ$, the benchmark value required for the Kozai 
effect to operate (equation [\ref{kozaicross}]).

[9] In addition to the ejection of planets during the scattering
encounters, orbital eccentricites can be increased so that planetary
orbits will cross each other. Most solar systems in such states will
eject --- or perhaps accrete --- planets on relatively short time
scales. For systems with the architecture of the current solar system,
the cross section for this channel of secondary ejection is comparable
to that of direction ejection, so that the total cross section for
ejection is effectively doubled (Figure \ref{fig:eject}). For the
ultra-compact configuration of the solar system (in or near multiple
mean motion resonances), the cross section for ejection due to orbit
crossing is comparable to that of the standard solar system, but the
cross section for direct ejection is smaller.

[10] The cross sections for changing the semimajor axes of the
planetary orbits are smaller than those for increasing eccentricity
and/or inclination angle (Figure \ref{fig:across}). Equivalently, the
semimajor axes change much less than the other orbital elements during
scattering encounters. In rough terms, 10\% changes in the semimajor
axis --- for planets that remain bound --- have approximately the same
cross sections as planetary ejection.

[11] The cross sections for solar systems interacting with single
stars are smaller than those for binary encounters (Figure
\ref{fig:singlecross}). The single-star cross sections are nearly
self-similar (Figure \ref{fig:singlescale}), and scale with the
semimajor axis of the planet and cluster velocity dispersion according
to $\cross\propto{a}\,\,\vdisp^{6/5}$. The scaling exponent for
velocity is somewhat smaller than that for binaries and the dependence
of the cross sections on the post-encounter eccentricity is steeper.
On average, the single-star cross sections are smaller than the binary
cross sections by a factor of $\sim3.6$ for small velocity dispersions
($\vdisp=1$ km/s). This factor falls to only $\sim2$ for larger values
$\vdisp\sim20$ km/s; for higher speeds we expect the binary components
to act as two separate stars during the encounters (except for close
binaries). In general, the effective cross section is a linear 
combination of the single and binary star cross sections, 
\be
\cross = f_b \cross_{\rm binary} + (1-f_b) 
\cross_{\rm single}\,,
\ee
where $f_b$ is the binary fraction. 

[12] We have briefly revisited the dynamical constraint that can be
placed on the birth aggregate of our solar system due to scattering
encounters (Section \ref{sec:application}).  The strength of this
constraint depends crucially on whether one assumes that the solar
system forms in a robust, long-lived cluster (with $\tau \gta 100$
Myr, like those that become open clusters) or in a short-lived cluster
that dissipates within $\tau\sim10$ Myr. For long-lived clusters, the
requirement that the solar system is not disrupted implies an order of
magnitude upper limit on the solar birth aggregate of $N \lta 10^4$
(see equation [\ref{constraint}]). In practice, one should construct
the probability distribution for solar system survival/disruption as a
function of $N$ (using the cross sections determined herein), and
combine it with the other constraints on the birth cluster (see Figure
7 in \citealt{adams2010}; see also \citealt{zwart} and
\citealt{pfalzner}).

[13] The cross section for removing a solar system from mean motion
resonance is much higher than that required to disrupt the planetary
orbits. For the ultra-compact multi-resonant configuration advocated
by some versions of the Nice model, this cross section (see equation
[\ref{rescross}]) is $\sim9$ times larger than the disruption cross
section for the usual solar system architecture. If removal from
resonance leads to longer-term instability, then constraints on the
solar birth aggregate would be tighter for systems in the
multi-resonant configuration.

\subsection{Discussion} 
\label{sec:discuss} 

The cross sections reported in this paper are subject to three
different types of uncertainties, and the distinctions among these
quantities should be kept in mind. [1] First, the Monte Carlo
procedure used to determine specific cross sections (as outlined in
Section \ref{sec:formulate}) results in uncertainties due to incomplete
sampling. These uncertainties decrease with increasing size of the
ensemble of simulations and are proportional to ${\cal N}_E^{-1/2}$.
Over most of the parameter space, we run sufficient numbers ${\cal
  N}_E$ of scattering experiments so that the sampling errors are less
than $\sim5\%$ and usually even smaller. These sampling errors are
present in all of the cross sections presented in Section
\ref{sec:results}, although they are usually not included on the plots
(however, see Figure \ref{fig:efinite}).  [2] Next, in Section
\ref{sec:analysis}, we explore scaling laws to collapse the cross
sections for varying velocity dispersion $\vdisp$, host mass $M_\ast$,
and planet semimajor axis $a$ into nearly self-similar forms. The
range of the resulting scaled functions is thus characterized by the
error bars shown in Figures \ref{fig:scale}, \ref{fig:vscale},
\ref{fig:mscale}, \ref{fig:iscale}, and \ref{fig:singlescale}. These
error bars represent a measure of the degree to which the cross
sections depart from self-similarity. The size of these error bars
falls in the range 10 -- 15\%, except for the scaling with the mass of
the host star (where the error bars correspond to 20\% departures).
[3] Finally, the mean of the scaled cross sections are described by
fitted functions with simple forms. The differences between these
functions and the mean scaled cross sections are of order 5 -- 10\%,
smaller than the standard deviations of the different sets of cross
sections used to construct the mean forms. 

In addition to the uncertainties outlined above, the cross sections
calculated herein depend on the features of the stellar population
that provides the perturbations. As described in Section
\ref{sec:formulate}, the cross sections sample the distributions of
stellar masses, binary periods, binary mass ratios, binary orbital
eccentricities, etc. Different choices for these distributions will
lead to corresponding variations in the cross sections. Although we
use observations to specify the distributions, they are nonetheless
subject to both measurement error and possible variations from region
to region.

The numerical simulations carried out for this paper determine the
immediate changes in the orbital elements of the solar systems due to
passing stars. However, additional changes in the orbital elements can
occur over longer time scales. As one example, after an encounter, a
planetary system often has larger eccentricities, which can lead to
orbital instability over longer spans of time. But the timescales for
such instabilities can have a wide range. For systems where the
eccentricities are increased so much that planetary orbits cross, one
expects instability and (usually) planet ejection on a relatively
short time. The cross sections for orbit crossing are thus of great
interest and are given in Figure \ref{fig:eject}. For systems with
smaller eccenticity increases, however, orbital instability can take
much longer. For compact multi-resonant solar systems, modest changes
in the orbital elements and/or the removal of the system from its
resonant state can lead to instabilities over millions of years
\citep{batbrown}. For systems with more widely separated orbits,
instabilities can take even longer than the current age of the
universe \citep{batlaugh,laskar}. To study this issue, the
post-encounter solar systems must be integrated over long time scales
(up to billions of years) to fully determine the effects of the
encounters. This task is left for future work. On another front, the
orbits could also damped after the scattering encounters, thereby
moving the orbits back towards smaller eccentricities
\citep{levison2007,picogna}. This effect should also be considered in
follow-up studies, especially on time scales of 1 -- 100 Myr when
solar systems are expected to retain a significant population of
planetesimals.

The scattering encounters considered herein can be effective in
sculpting giant planet orbits and the Kuiper Belt of our Solar System
(e.g., \citealt{kenyon}). On the other hand, the Oort cloud is too
large to be produced within a young embedded cluster (e.g., see
\citealt{brasser2012} for further discussion). More specifically, the
Oort cloud extends out to $\sim50,000$ AU \citep{oort,jewitt}, more
than 1000 times the size of the solar systems considered in this
paper.  With this enormous size, the Oort cloud would be decimated by
passing stars within the cluster.  As a result, the cloud must be
produced later, after the solar system leaves its birth cluster, or
perhaps during its exit. Any viable scenario for the solar birth
environment must simultaneously account for the Oort cloud, the giant
planet orbits, Kuiper Belt properties, radioactive enrichment, Sedna's
orbit, and survival of the solar nebula gas reservoir; these coupled 
constraints thus pose an interesting and challenging opimization
problem for further study.

\medskip 
\textbf{Acknowledgments:} We would like to thank Konstantin Batygin,
Greg Laughlin, Fred Rasio, and Maxwell Tsai both for early motivation
and subsequent useful discussions. This collaboration was initiated
through the 2014 International Summer Institute for Modeling in
Astrophysics (ISIMA), which focused on gravitational dynamics, and was
hosted by the Canadian Institute for Theoretical Astrophyics (CITA).
The numerical calculations were performed at the Harvard Smithsonian
Center for Astrophysics (CfA), on a cluster supported by the Institute
for Theory and Computation (ITC). We are grateful for the hospitality
and resources of CfA, CITA, ISIMA, ITC, and Univ. Michigan.

\label{lastpage}  
 
\end{document}